\documentclass[preprint,12pt,authoryear]{elsarticle}
\usepackage[utf8]{inputenc}
\usepackage{geometry}
\usepackage{graphicx}
\usepackage{amsmath, amssymb}
\usepackage{booktabs} 
\usepackage{hyperref}
\usepackage{tikz}
\usepackage{multirow} 
\usepackage{algorithm}
\usepackage{algorithmic}
\usepackage{threeparttable}
\usepackage{subcaption}
\usepackage{mathtools}
\usepackage{amsthm}
\usepackage{cleveref}
\usepackage{pifont}
\usepackage{enumitem}
\usepackage[left]{lineno}

\journal{Information Processing \& Management}
\begin{document}

\begin{frontmatter}
\title{S2-CAR: Segmentation-Supervised Complexity-Adaptive Recommendation}

\author[1]{Linjiang Guo}
\author[1]{Nitin Bisht}
\author[2]{Shiqing Wu}
\author[1]{Xianzhi Wang}
\author[3]{Guandong Xu\corref{cor1}}

\cortext[cor1]{Corresponding author. Email: gdxu@eduhk.hk}

\affiliation[1]{
    organization={University of Technology Sydney},
    addressline={15 Broadway},
    city={Ultimo},
    postcode={2007},
    state={NSW},
    country={Australia}
}
\affiliation[2]{
    organization={City University of Macau},
    addressline={Avenida Padre Tomás Pereira},
    city={Taipa},
    country={Macau SAR, China}
}
\affiliation[3]{
    organization={The Education University of Hong Kong},
    addressline={10 Lo Ping Road},
    city={Tai Po},
    state={New Territories},
    country={Hong Kong SAR, China}
}

\ead{gdxu@eduhk.hk}

\begin{abstract}
Sequential recommendation aims to predict user preferences from interaction histories, yet existing models often struggle when behavior patterns become complex and heterogeneous. A key reason is that interaction histories are rarely uniform: users' interests shift in a latent way over time, yet existing models either treat the full sequence as a homogeneous context or rely on rigid time-window segmentation that misaligns with true intent boundaries. This mis-segmentation not only introduces cross-intent interference at intermediate sequence positions but also leads to over-reliance on short-term interest signals. To address this, we propose \textbf{S2-CAR}, a segmentation-supervised and complexity-adaptive framework for sequential recommendation that models user intent as a continuous latent energy state. Specifically, it uses the Context-Aware Soft Temporal Point Process (Soft-TPP) to segment boundaries triggered by the natural decay of latent-state energy rather than fixed intervals, enabling intent segmentation without fixed time-gap rules. Next, upon this segmentation, a Segment-Count-Adaptive Multi-Intent Extraction module hierarchically aggregates intent-coherent segments into a compact set of multi-interest representations. Extensive experiments on 3 representative public benchmark datasets spanning movie, e-commerce, and gaming domains across 13 baselines demonstrate that S2-CAR consistently outperforms state-of-the-art methods across all datasets and metrics. Further analysis shows that the proposed energy-based segmentation serves as a plug-and-play module, yielding consistent improvements when integrated into existing sequential recommendation backbones. 
\end{abstract}


\end{frontmatter}
\resetlinenumber[1]

\section{Introduction}

\begin{figure}[htbp]
    \centering
    \begin{subfigure}[t]{0.49\textwidth}
        \centering
        \includegraphics[width=\textwidth]{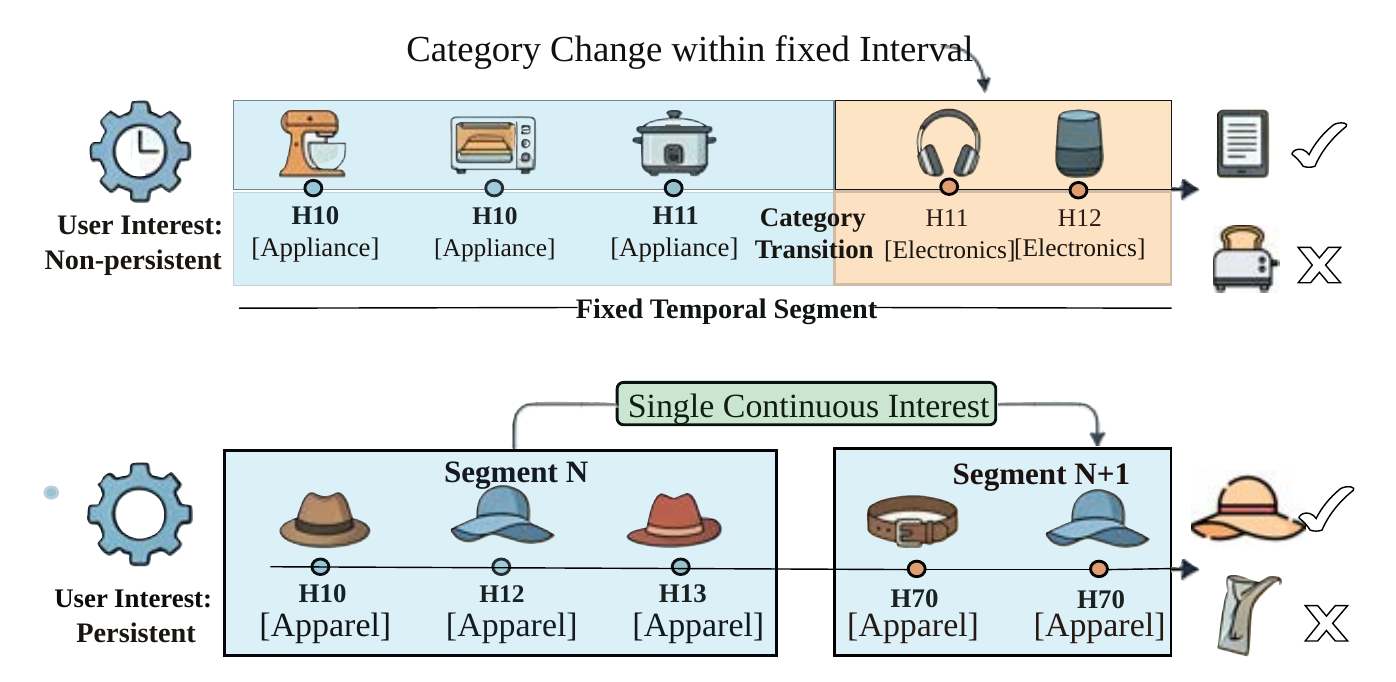}
        \caption{Two toy failure modes of fixed temporal segmentation: missed intent transitions within a segment and artificial splitting of a persistent interest across segments. H denotes the timestamp of interactions (in hours).}
        \label{fig:left}
    \end{subfigure}
     \hfill
    \begin{subfigure}[t]{0.49\textwidth}
        \centering
        \includegraphics[width=\textwidth]{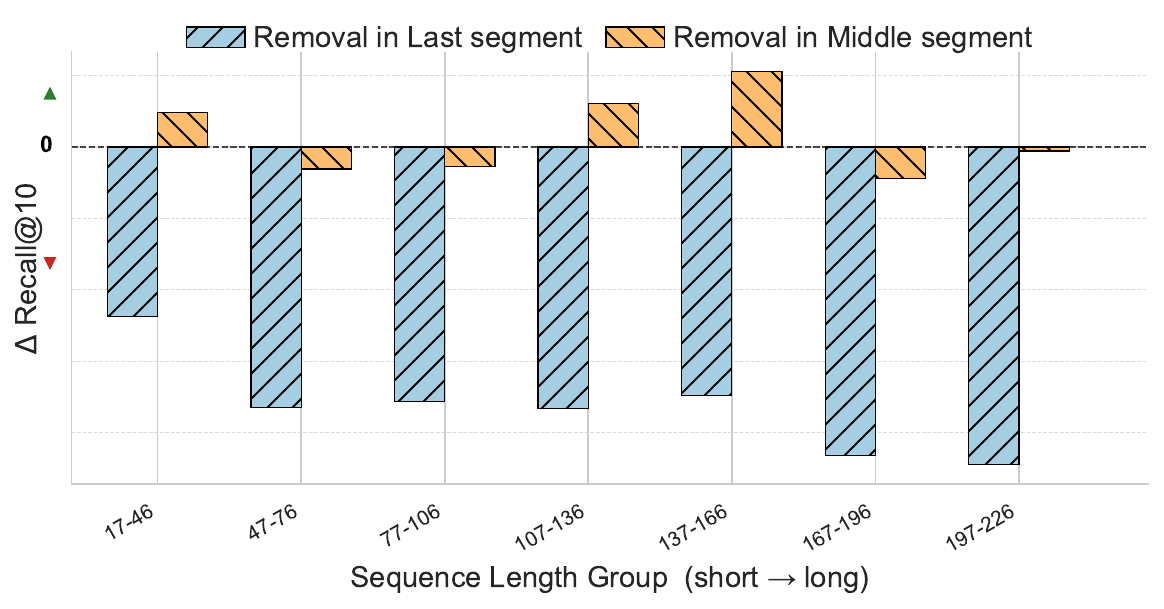}
        \caption{Performance impact of random item deletion from the last segment (short-term interest proxy) and middle segment (long-term interest proxy) across sequence length groups, measured by $\Delta$ Recall@10 (deleted sequence minus original sequence; positive values indicate improvement after deletion, and negative values indicate degradation).}
        \label{fig:right}
    \end{subfigure}
    \caption{Limitations of fixed temporal segmentation in sequential recommendation.}
    \label{fig:both}
\end{figure}

Sequential recommendation aims to predict a user's next interaction by modeling their behavior history~\citep{fang2020deep, li2021hyperbolic, peng2025tagrec}. Modeling long behavior sequences has become increasingly important for capturing both recent interests and long-range preferences, given the rapid growth of user activity on modern platforms. Existing sequential recommendation models, such as recurrent architectures (e.g., GRU4Rec~\citep{hidasi2015session}) and self-attention-based methods (e.g., SASRec~\citep{kang2018self} and BERT4Rec~\citep{sun2019bert4rec}), have achieved strong performance by learning item-transition patterns from historical interactions. Despite their effectiveness, they often struggle when sequences become long and behavior patterns become increasingly complex. As illustrated in \Cref{fig:left}, a key reason is that long interaction histories are rarely homogeneous, with users' interests often shifting rapidly, leading to a noisy collaborative signal. Conversely, user preferences sometimes persist across temporal sessions, resulting in context dilution. Under such scenarios,  existing models can blur the distinction between genuine recent intent and irrelevant or repeated historical signals. 

To empirically validate this problem, we conduct an empirical study using SASRec as a representative sequential backbone, in which we randomly remove half of the items from either the last segment (as a proxy for short-term interest) or the middle segment (as a proxy for long-term interest) of each user's interaction history, and evaluate the resulting change in sequential recommendation performance in terms of recommendation metric, i.e., Recall@10. Interestingly, as illustrated in \Cref{fig:right}, two key observations emerge. First, for a subset of users, deleting items from the middle segment does not degrade performance; instead, it even improves it. This observation suggests a phenomenon: existing models are insensitive to middle-segment content, and this insensitivity becomes more pronounced as sequence length increases. Whether this stems from cross-intent noise accumulated in the middle portion, or from boundary definition errors inherent in the fixed segmentation used to construct the experiment, both explanations converge on the same underlying failure: existing models systematically under-utilize interaction signals embedded in the middle of long sequences. {\color{black}We later revisit this protocol in \Cref{sec:exp4} to evaluate whether the proposed model resolves this limitation.} Second, deleting items from the last segment results in consistent, increasingly severe performance degradation as sequence length increases, indicating that existing models remain heavily reliant on short-term signals. Together, these findings suggest that the core challenge in sequential recommendation is not merely the attenuation of early interactions over time, but the absence of a principled mechanism for identifying genuine intent boundaries, without which models cannot reliably distinguish informative historical signals from irrelevant or redundant ones.

Recently, some papers have tried to mitigate this issue. Hierarchical methods \citep{quadrana2017personalizing,sharma2024survey,zhang2025hierarchical} attempt to partition the interaction sequence into sub-sessions before modeling. Similarly, segmentation strategies typically rely on fixed (e.g., 30-minute) gap-based \citep{ludewig2018evaluation,wang2021survey}, token-similarity-based fragment reorganization \citep{li2024item}, or multi-intent projection \citep{choi2024multi} and aggregation mechanisms~\citep{long2025dual}. However, these approaches share a common limitation: their segmentation criteria are largely heuristic and are defined independently of the user's actual behavioral dynamics. In the real world, such strategies frequently fail because they conflate temporal distance with co-occurrence divergence, causing asynchronous intent shifts to be mishandled. For example, a user may revisit the same intent after a long pause or switch interests within a short session. Additionally, existing multi-interest modules usually allocate a uniform slot capacity to all segments, ignoring that segments can differ substantially in temporal span and internal uncertainty.

To solve this, we propose our novel framework: Segmentation-Supervised Hierarchical Multi-Intent Network with Complexity-Adaptive Regularization \textbf{S2-CAR}. However, we face the following technical challenges: (1) how to infer intent boundaries from behavioral dynamics rather than fixed temporal-window rules; (2) how to compactly represent recurring and heterogeneous interests without allocating redundant capacity; and (3) how to regularize sequential representations with segment-level structure without sacrificing inference-time causal fidelity. To answer this, we first partition the continuous interaction stream into behaviorally coherent segments. We introduce a Context-Aware Soft Temporal Point Process that models the user's latent intent as a continuous-time energy state decaying between interactions. A segment boundary is triggered only when the energy retention ratio falls below a dataset-specific threshold, resulting in task-specific intent segments without manually annotated boundary labels. Next, to compress $K$ intent segments into a compact set of interest slots that adapts to segment complexity, we propose a Segment-Count-Adaptive Multi-Intent Extraction module, where a compression ratio $\rho$ consolidates cross-segment redundancy caused by periodic behavioral recurrence and prevents capacity waste in the interest representation. Finally, to bridge the structural gap between fine-grained sequential representations and globally balanced multi-interest representations, we design a Hierarchical Interest Encoder that serves as a contrastive supervision path, aligning the two complementary views via a contrastive loss without degrading inference-time sequential fidelity.

The main contributions of this paper are summarized as follows:
\begin{itemize}
    \item We propose a Context-Aware Soft-TPP for unsupervised energy-based intent segmentation. The resulting boundary flags guide downstream modules without requiring manually annotated boundary labels, replacing rigid fixed-window partitioning strategies while remaining fully compatible with end-to-end training pipelines.
    
    \item We introduce a Segment-Count-Adaptive Multi-Intent Extraction mechanism that adaptively gates the number of active interest slots according to segment count and consolidates periodic cross-segment redundancy using a validation-selected compression ratio $\rho$.
    
    \item We further design a decoupled Hierarchical Interest Encoder that provides structurally balanced supervision during training and regularizes the Sequential Inference Encoder via contrastive loss.
    
    \item We conduct extensive experiments on three representative public benchmark datasets (movie, e-commerce, and gaming domains) against 13 baselines, demonstrating that S2-CAR consistently achieves state-of-the-art performance, with particularly significant gains for users with long and complex interaction histories.
\end{itemize}

\section{Related Work}

\subsection{Sequential Recommendation}
Sequential recommendation aims to model the dynamic evolution of user
preferences by leveraging chronologically ordered interaction histories to
predict the next item of interest~\citep{wang2021survey}.
Early work in this area employed recurrent neural networks to capture
temporal dependencies: GRU4Rec~\citep{hidasi2015session} introduced
session-parallel mini-batching with gated recurrent units, while
Caser~\citep{tang2018personalized} reframed the problem as image recognition
by applying convolutional filters over a sliding window of recent
item embeddings to capture short-range transition patterns.
The advent of the Transformer architecture brought a significant
paradigm shift. SASRec~\citep{kang2018self} demonstrated that a lightweight
unidirectional self-attention mechanism could outperform recurrent models
by adaptively attending to relevant items across the full sequence.
BERT4Rec~\citep{sun2019bert4rec} extended this line by adopting bidirectional
attention with a cloze-style masked item prediction objective, enriching
item representations with both left- and right-context signals.

Despite their strong empirical performance on standard benchmarks, these
methods share a weakness when applied to long interaction sequences. The autoregressive aggregation chain inherent to both recurrent and attention-based encoders progressively attenuates the influence of early interactions, causing relevant historical interactions to be buried in the middle of long sequences. This issue is analogous to the 'Lost-in-the-Middle' phenomenon observed in large language
models~\citep{liu2024lost,zhang2024found}, where critical context positioned in the
middle of a long input tends to be systematically under-utilized by
autoregressive decoders.

To alleviate the resulting noise and intent ambiguity in long interaction
sequences, several methods introduce session-based segmentation as a
preprocessing step. TLSRec~\citep{chen2022time} adopts a fixed
time-gap threshold to split sequences into sub-sessions, while
other approaches apply sliding windows of fixed length~\citep{tang2018personalized,qin2024intent,wang2025intent}.
However, these segmentation criteria are defined independently of the
user's actual behavioral dynamics: fixed windows are insensitive to
asynchronous intent shifts, and raw time-gap thresholds conflate temporal
distance with co-occurrence divergence, as a user may revisit the same
intent after a prolonged pause or transition to a new interest within a
single short session.

\subsection{Multi-Interest Recommendation}
A single vector representation is often insufficient to capture the
heterogeneous and evolving nature of user preferences, motivating a line
of research on multi-interest modeling~\citep{deng2024sse4rec,li2025multi,lv2025dynamic}.
Early work in this direction, such as MIND~\citep{li2019multi} and
ComiRec~\citep{cen2020controllable}, introduced capsule routing and
multi-head attention mechanisms, respectively, to extract a fixed set
of interest vectors from the full interaction history, where each vector encodes a stable, recurring preference pattern of the user.
SINE~\citep{tan2021sparse} further improved scalability by sparsely
activating only a subset of conceptual prototypes for each user,
reducing the redundancy inherent in dense multi-interest representations.
DSSRec~\citep{ma2020disentangled} introduces latent self-supervision to
disentangle multiple behavioral factors from the interaction sequence,
enabling the model to separately represent and recombine independent user
motivations.
MiaSRec~\citep{choi2024multi} augments the standard Transformer encoder
with a multi-intent module that routes each item to  several latent embeddings, allowing concurrent tracking of
multiple evolving interests within a unified architecture.

A related but distinct line of work focuses on modeling short-term user
intent, i.e., the immediate behavioral goal driving a specific
interaction session, rather than long-term interest patterns.
NOVA~\citep{liu2021noninvasive} introduces non-invasive self-supervised
signals to capture intent-level transitions without modifying the
primary sequence encoder.
ICSRec~\citep{qin2024intent} decomposes interaction histories into
intent-homogeneous subsequences through item clustering and applies
contrastive learning at the intent level to explicitly align
representations of sequences sharing the same underlying user purpose.

Despite operating at different granularities, these approaches share two
common limitations.
First, they allocate a uniform number of interest slots across all users
and sessions, ignoring the substantial variability in temporal span and
behavioral complexity across different interaction phases.
Next, they do not account for periodic intent recurrence, wherein the
same short-term goal manifests repeatedly across non-adjacent sessions;
without explicit consolidation, the same underlying interest is
wastefully re-allocated to multiple redundant slots rather than being
unified into a single compact representation.

\subsection{Temporal Point Process}
Temporal Point Processes (TPPs) provide a principled probabilistic
framework for modeling the occurrence of discrete events in continuous
time~\citep{daley2003introduction}.
Classical TPP models, including the Hawkes process~\citep{hawkes1971spectra}
and its variants, such as the self-correcting process~\citep{isham1979self}
and the mutually exciting multivariate Hawkes process,
capture self-exciting or inhibitory dynamics through parametric intensity
kernels, but rely on functional forms that limit their
expressiveness in complex real-world settings.

Neural TPP models overcome this limitation by parameterizing the intensity
function with recurrent or attention-based architectures.
RMTPP~\citep{du2016recurrent} first demonstrated that a recurrent neural
network could jointly model event types and inter-event times by encoding
the full event history into a continuous hidden state.
NHP~\citep{mei2017neural} extended this line with a continuous-time LSTM
that evolves the hidden state between events according to an exponential
decay, enabling more faithful modeling of the inter-event dynamics.
Recent work further replaced recurrence with Neural TPP models, achieving strong performance on event streams such as financial
transactions~\citep{du2016recurrent}, social network
activities~\citep{zhao2015seismic}, and electronic health
records~\citep{mei2017neural}.

In the context of user behavior modeling, TPPs have been applied primarily
to next-event time prediction, estimating when a user will next interact
with the system. Coevolving method~\citep{wang2016coevolutionary} models the
co-evolution of user and item states in continuous time using coupled
point processes, capturing the mutual influence between user engagement
and item popularity dynamics. THRNN~\citep{vassoy2019time} introduces a joint hierarchical RNN 
and a point process model that jointly models session-level and 
event-level dynamics, providing a closer connection to the session 
structure inherent in recommendation settings.
TLSRec~\citep{chen2022time} incorporates inter-event intervals as auxiliary
signals to modulate attention weights in a sequential encoder, improving
sensitivity to temporal irregularity. However, it treats temporal
information purely as a feature rather than exploiting it as a generative
mechanism for inferring latent intent structure.
Our proposed method S2-CAR introduces a Soft-TPP that models the user's latent
intent as a scalar retention ratio governed by a context-adaptive ordinary
differential equation; a segment boundary is triggered when the energy
retention ratio drops below a dataset-specific threshold, yielding task-specific
intent segments without manually annotated boundary labels or fixed time-gap rules.

\section{Methodology}

In this section, we introduce our novel framework \textbf{(S2-CAR)}, a Segmentation-Supervised Hierarchical Multi-Intent Network with Complexity-Adaptive Regularization.
We begin by introducing the notations used in the paper, which are summarized in~\Cref{tab:notation}. Let $\mathcal{U}$ and $\mathcal{V}$ denote the sets of users and items, respectively. For a user $u$, the interaction history is a chronologically ordered sequence $S = \{(v_1, c_1, t_1), (v_2, c_2, t_2), \dots, (v_N, c_N, t_N)\}$, where $v_i \in \mathcal{V}$ is the interacted item, $c_i$ is its category, and $t_i$ is the timestamp. The task is to predict the next item $v_{N+1}$ given the sequence $S$.

\subsection{Framework Overview}

\begin{figure}[tbp]
    \centering
    \includegraphics[width=\linewidth]{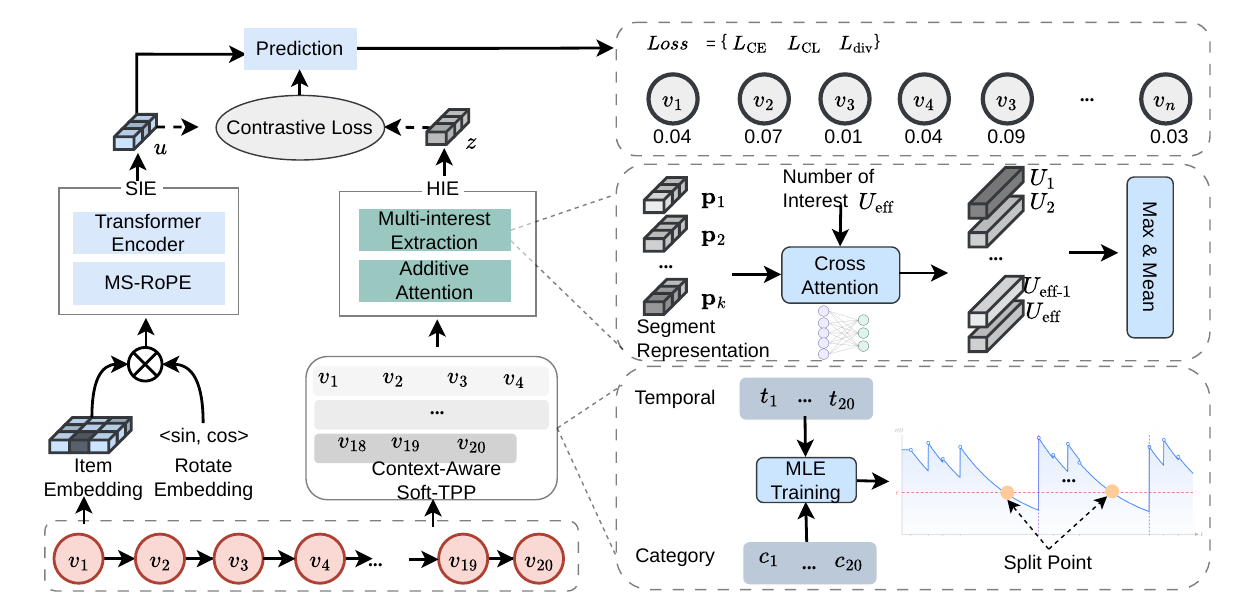}
    \caption{
        Overview of our proposed framework. 
        A frozen Soft-TPP model produces binary boundary flags from the raw interaction sequence. 
        The Sequential Inference Encoder uses a boundary-aware causal Transformer 
        with MS-RoPE to produce the user embedding $\mathbf{u}$for next-item scoring. 
        The Hierarchical Interest Encoder builds segment-level 
        multi-interest representations and provides a contrastive regularization signal.
    }
    \label{fig:framework}
\end{figure}
A challenge in sequential recommendation is that a user's intent is not static, as it shifts across different temporal phases of their interaction history. Existing methods either ignore these shifts entirely or rely on heuristic-selected fixed-length windows or raw time-gaps to partition the sequences~\citep{chen2022time,qin2024intent}. While fixed windows are insensitive to actual behavioral transitions, raw time-gaps are a poor proxy for intent boundaries (i.e., a user may return to the same interest after a long pause, or shift to an entirely different one within a single short session).

To solve this, we treat intent as a continuous latent state that decays between interactions. A boundary is considered when the residual energy of the latent state drops below a threshold, indicating a genuine intent transition. This segmentation is learned using a temporal point process objective and then frozen, such that downstream modules always receive behaviorally coherent segments rather than arbitrary chunks. The overall framework (as illustrated in \Cref{fig:framework}) contains a Sequential Inference Encoder and a Hierarchical Interest Encoder branch that share the same boundary-aware token representations but serve distinct roles. Firstly, a Context-Aware Soft Temporal Point Process (abbreviated as Soft-TPP hereafter) model is pretrained to capture the continuous-time dynamics of user intent. After pretraining, it is frozen and used to produce binary boundary flags (\Cref{sec:tpp}). The Sequential Inference Encoder  (\Cref{sec:bottom}) consists of a boundary-aware causal Transformer with Multi-Scale Rotary Position Embedding (MS-RoPE) that encodes the full item sequence and extracts the user embedding $\mathbf{u}$from the last valid position. This embedding is directly used for next-item scoring. The Hierarchical Interest Encoder (\Cref{sec:segment}) is an auxiliary training branch built on top of the SIE hidden states: it uses the frozen TPP boundary flags to aggregate token-level representations into segment-level representations, extracts multi-interest vectors via cross-attention, and provides a contrastive supervision signal during training.

\subsection{Context-Aware Soft-TPP}
\label{sec:tpp}

\subsubsection{Continuous-Time Intent Dynamics}

To capture how user intent evolves between consecutive interactions, we model the latent intent as a d-dimensional energy vector $\mathbf{e}(t) \in \mathbb{R}^d$. The scalar event intensity $\lambda(t)$ is defined as the squared $\ell_2$ norm of this vector, i.e., $\lambda(t) = |\mathbf{e}(t)|_2^2$. For boundary detection, we further derive a scalar retention ratio $r_i$ by averaging the per-dimension exponential retention values. Intuitively, $\mathbf{e}(t)$ reflects how strongly the current intent remains active over time. We define the latent state as:
\begin{equation}
\mathbf{e}(t) = \exp\!\left(-\boldsymbol{\delta}_i (t - t_i)\right),
\label{eq:ode_solution}
\end{equation}
where $\boldsymbol{\delta}_i \in \mathbb{R}_{+}^d$ is the dimension-wise decay
rate, and the exponential is applied element-wise. At each event position $i$, the energy vector is reset to an all-ones vector $\mathbf{e}(t_i) = \mathbf{1_d}$, modeling the assumption that a new interaction fully reactivates the current intent state before it begins to decay.
Equivalently, $\mathbf{e}(t)$ satisfies the following ordinary differential equation:
\begin{equation}
\frac{d \mathbf{e}(t)}{dt} = -\boldsymbol{\delta}_i \odot \mathbf{e}(t),
\quad t \in (t_i,\, t_{i+1}),
\label{eq:ode}
\end{equation}

\subsubsection{Context-Adaptive Decay}

A key challenge in modeling real-world interaction sequences is that user
intent does not shift at a uniform rate. Interactions exhibiting strong
co-occurrence patterns should sustain the current latent state, whereas
categorical divergence should accelerate its dissipation. A fixed decay rate
cannot capture this variability. Moreover, the same local category transition
may carry different semantic weight depending on the user's long-term
behavioral profile: a shift from Electronics to Appliances signals a stronger
intent boundary for a user whose history is concentrated in a single domain
than for one who habitually browses across diverse categories. We therefore
introduce a Context-Adaptive Decay mechanism, where the dimension-wise decay
rate $\boldsymbol{\delta}_i$ is conditioned on the local category transition
$(c_{i-1}, c_i)$ and a user-level context vector $\bar{\mathbf{h}}^c_i$ that
summarizes the user's historical category distribution up to position $i$:
\begin{equation}
\begin{split}
\boldsymbol{\delta}_i &= \varphi\!\left(
    f_\delta\!\left(\left[
        \mathbf{h}^{c}_i \,\|\, \mathbf{h}^{c}_{i-1} \,\|\,
        \bar{\mathbf{h}}^{c}_i
    \right]\right)
\right) + \epsilon, \\
\bar{\mathbf{h}}^{c}_i &= \frac{1}{i}\sum_{j=1}^{i} \mathbf{h}^{c}_j,
\end{split}
\label{eq:delta}
\end{equation}
where $\mathbf{h}^{c}_i, \mathbf{h}^{c}_{i-1} \in \mathbb{R}^d$ are the
category embeddings of the $i$-th and $(i{-}1)$-th interactions obtained
from category embeddings $\mathbf{E}_C$. For the first position, we set
$\mathbf{h}^{c}_0=\mathbf{0}$ so that the initial transition context is
well-defined without using any future information.
$\bar{\mathbf{h}}^{c}_i$ is the causal running mean of category embeddings
up to position $i$, $[\,\cdot\,\|\,\cdot\,]$ denotes vector concatenation.
$\varphi$ is the Softplus activation.
$f_\delta : \mathbb{R}^{3d} \to \mathbb{R}^d$ is a MLP, and
$\epsilon = 10^{-6}$ ensures strict positivity.
The local transition pair $(\mathbf{h}^{c}_{i-1}, \mathbf{h}^{c}_i)$
provides a sharp, position-specific signal of categorical divergence,
while $\bar{\mathbf{h}}^{c}_i$ serves as a user-level modulator encoding
the user's long-term category distribution.

\subsubsection{Soft-TPP Formulation}

We initialize the Temporal Point Process framework with two design choices
tailored to intent modeling. The event intensity is defined as the aggregate
squared residual energy across all latent dimensions:
\begin{equation}
\lambda(t) = \sum_{k=1}^{d} e^{-2\delta_{i,k}(t - t_i)}
\label{eq:intensity}
\end{equation}
where the per-dimension energy at time $t$ is
$e_k(t) = e^{-\delta_{i,k}(t-t_i)}$ (Eq.~\eqref{eq:ode_solution}), and $\delta_{i,k}$ denotes the $k$-th element of the decay vector $\boldsymbol{\delta}_i$.
The intensity can equivalently be written as
$\lambda(t) = \sum_{k=1}^{d} e_k(t)^2$,
i.e.,\ the squared $\ell_2$ norm of the latent energy vector.
This choice is motivated by two considerations.
First, the quadratic decay rate makes $\lambda(t)$ more sensitive to
energy dissipation than a linear sum $\sum_k e_k(t)$ would be,
producing sharper boundary signals when the latent state approaches
exhaustion.
Second, it results in a compensator that enables efficient computation without numerical approximation:
\begin{equation}
\Lambda_i = \int_{t_i}^{t_{i+1}}\lambda(t)\,\mathrm{d}t
 = \sum_{k=1}^{d}\frac{1}{2\delta_{i,k}}
   \bigl(1 - e^{-2\delta_{i,k}\Delta t_i}\bigr),
\label{eq:compensator}
\end{equation}

This formulation encodes the intuition that a high-energy latent state
indicates an active intent, making future interactions more likely, while
a decayed state signals intent exhaustion. However, not all transitions
should contribute equally to the likelihood: a cross-category jump reflects
an intent shift rather than a continuation of the current trajectory.
We therefore introduce a soft co-occurrence weight $w(c_i, c_{i+1}) \in (0,1]$
that scales each event's \emph{complete} log-likelihood contribution
according to categorical relatedness:
\begin{equation}
w(c_i, c_{i+1}) = \begin{cases}
1 & \text{if } c_i = c_{i+1} \\
\sigma\!\left(\alpha \cdot \cos(\mathbf{h}^{c}_i,\,\mathbf{h}^{c}_{i+1})\right)
  & \text{otherwise}
\end{cases}
\label{eq:soft_weight}
\end{equation}
where $\alpha$ is a scaling factor and $\sigma(\cdot)$ is the sigmoid
function. This weighted objective can be viewed as a likelihood reweighting
scheme that adjusts the contribution of each observed transition while
preserving the TPP likelihood structure~\citep{ogata1978asymptotic,hu2002weighted}.
Intuitively, $w_i$ controls how much each observed transition contributes to
the weighted likelihood, while the direction of
the update for $\boldsymbol{\delta}_i$ is still determined by the TPP
likelihood and the context-adaptive decay network. The learned decay rates,
together with the elapsed time $\Delta t_i$, determine the energy retention
ratio $r_i$ used for boundary detection. Substituting
Eq.~\eqref{eq:soft_weight} into the standard MLE objective gives the
Soft-TPP loss:
\begin{equation}
\mathcal{L}_{\text{TPP}} = -\frac{1}{|\mathcal{P}|}\sum_{i \in \mathcal{P}}
w(c_i, c_{i+1})
\left[
  \log \lambda(t_{i+1})
  - \sum_{k=1}^{d} \frac{1}{2\delta_{i,k}}
    \left(1 - e^{-2\delta_{i,k}\Delta t_i}\right)
\right],
\label{eq:tpp_loss}
\end{equation}
where $\mathcal{P}$ denotes valid (non-padding) positions,
$\Delta t_i = t_{i+1} - t_i$,
$\lambda(t_{i+1}) = \sum_k e^{-2\delta_{i,k}\Delta t_i}$,
and the second term in the bracket is the compensator $\Lambda_i$
defined in Eq.~\eqref{eq:compensator}.

\subsubsection{Scale-Invariant Boundary Detection}
\label{sec:boundary}

After the Soft-TPP module is trained and frozen, it serves purely as a boundary detector. Since the latent intent state $\mathbf{e}(t) \in \mathbb{R}^d$ is a vector with per-dimension decay rates $\boldsymbol{\delta}_i \in \mathbb{R}^d$, we define the Energy Retention Ratio as the mean of per-dimension exponential decay:
\begin{equation}
r_i = \frac{1}{d}\sum_{k=1}^{d} \exp(-\delta_{i,k} \cdot \Delta t_i),
\label{eq:ratio}
\end{equation}
where $\Delta t_i = t_{i+1} - t_i$. 
A boundary is inserted before $v_{i+1}$ when $r_i < \tau$, where $\tau \in (0, 1)$ is a threshold hyperparameter. A lower $\tau$ requires more severe energy loss before declaring a new segment, yielding coarser segmentation, while a higher $\tau$ is more sensitive to minor intent shifts.

\subsubsection{Two-Stage Decoupled Training}

The TPP is trained in a fully unsupervised manner via maximum likelihood estimation (MLE) on the raw interaction history, without any manually annotated segment boundaries. All Soft-TPP parameters are frozen in their entirety before main model training begins. The recommendation model maintains its own item embedding table and does not share trainable parameters with the frozen Soft-TPP module. The TPP-derived boundary flags are used as fixed structural inputs, but no gradient from the downstream recommendation objective is propagated back to the Soft-TPP. This parameter and gradient isolation prevents the segmentation criterion from being directly optimized toward the next-item loss during main training, since the two objectives can conflict.

\subsection{Sequential Inference Encoder (SIE)}
\label{sec:bottom}

This path processes the full item sequence through a causal Transformer to produce the user embedding $\mathbf{u}$ used at inference. Boundary flags from the frozen TPP are injected as type embeddings, making the encoder structurally aware of intent transitions from the first layer without any architectural modifications to the attention mechanism.

\paragraph{Boundary Type Embedding}

The frozen Soft-TPP produces a binary boundary flag $b_i \in \{0, 1\}$ 
for each position $i$, where $b_i = 1$ marks the start of a new intent 
segment and $b_i = 0$ otherwise. These flags are fused into the item 
representation via a learnable boundary type embedding table 
$\mathbf{E}_{B} \in \mathbb{R}^{2 \times d}$:
\begin{equation}
\tilde{\mathbf{h}}_i = \mathbf{h}^{v}_i + \mathbf{E}_{B}(b_i),
\end{equation}
where $\mathbf{h}^{v}_i = \mathbf{E}_V(v_i) \in \mathbb{R}^d$ is the 
embedding of item $v_i$ looked up from the item embedding table 
$\mathbf{E}_V \in \mathbb{R}^{|\mathcal{V}| \times d}$, and 
$\mathbf{E}_{B}(b_i) \in \mathbb{R}^d$ is the corresponding 
boundary type vector, indicating whether the position is the boundary of the segment or not. This allows the encoder to be structurally aware 
of intent transitions from the first layer without any modification to 
the attention mechanism.

\paragraph{Multi-Scale Rotary Position Embedding}
The sequence $\{\tilde{\mathbf{h}}_i\}_{i=1}^N$ is processed by a stack
of $L_1$ Transformer blocks, where at each block the query
$\mathbf{q}_i^{(j)}$, key $\mathbf{k}_i^{(j)}$, and value $\mathbf{v}_i^{(j)}$ vectors for head $j$ are obtained via standard
linear projections of $\tilde{\mathbf{h}}_i$, each using multi-head self-attention with 
Multi-Scale Rotary Position Embedding.
Since user interaction sequences exhibit dependencies over varying time scales, from short-term co-occurrence patterns to long-term preference drift, a single frequency scale is insufficient to capture positional relationships across all granularities. For attention head $j$, the base frequency is scaled by $s_j = 10^{(j-1)/(H-1)}$ drawn from a log-uniform grid over $[1, 10]$:
\begin{equation}
\boldsymbol{\theta}_k^{(j)} = \frac{1}{s_j} \cdot \frac{1}{b^{2k/d_h}}, \quad
k = 1, \dots, d_h/2,
\end{equation}
where $b$ is the Rotary Position Embedding base and $d_h = d/H$ is the per-head dimension. Rather than adding position embeddings to the input, MS-RoPE implicitly encodes positional information by rotating the query and key vectors inside each attention head, which guarantees their inner product only depends on the relative distance rather than absolute positions, thereby eliminating the positional bias in attention score computation~\citep{su2024roformer}. The sequence $\{\tilde{\mathbf{h}}_i\}_{i=1}^N$, stacked as
$\mathbf{X} \in \mathbb{R}^{N \times d}$, is processed by Transformer blocks. The attention output for head $j$ is:

\newcommand{\Rij}[1]{\mathbf{R}_{#1}^{(j)}}
\begin{equation}
\begin{aligned}
\mathrm{Attn}^{(j)}(\mathbf{X}) &= \mathrm{softmax}\!\left(
\frac{\bigl(\Rij{i}\,\mathbf{q}_i^{(j)}\bigr)^\top
      \bigl(\Rij{l}\,\mathbf{k}_l^{(j)}\bigr) + \mathbf{C}_{il}}
{\sqrt{d_h}}\right)\mathbf{V}^{(j)}, \\
\bigl[\Rij{p}\,\mathbf{x}\bigr]_{2k-1:2k} &=
\begin{pmatrix}
\cos p\theta_k^{(j)} & -\sin p\theta_k^{(j)} \\
\sin p\theta_k^{(j)} &  \cos p\theta_k^{(j)}
\end{pmatrix}
\begin{pmatrix} x_{2k-1} \\ x_{2k} \end{pmatrix},
\quad k = 1,\dots,d_h/2,
\end{aligned}
\end{equation}
where $\mathbf{C}_{il} = -\infty$ if $i < l$ is the causal mask. By assigning a distinct frequency grid $\{\boldsymbol{\theta}_k^{(j)}\}$ to each head, MS-RoPE allows different heads to specialize in positional dependencies at different granularities from fine-grained short-term co-occurrence to coarse-grained long-term preference drift. Each block further applies a standard residual connection and position-wise feed-forward network. The layer-normalized output of the final block is $\mathbf{H} = [\mathbf{h}_1, \dots, \mathbf{h}_N] \in \mathbb{R}^{N \times d}$, shared by all downstream modules. The hidden state at the last position is taken as the user's sequential preference representation:
$\mathbf{u} = \mathbf{h}_N$.

\subsection{Hierarchical Interest Encoder (HIE)}
\label{sec:segment}

The Sequential Inference Encoder's causal readout at the last position $\mathbf{h}_N$ captures fine-grained 
sequential dependencies but carries a structural bias: early segments exert 
progressively attenuated influence through the autoregressive aggregation chain. The Hierarchical Interest Encoder 
addresses this complementary weakness by constructing a segment-level representation in 
which each intent phase contributes equally, independent of its temporal distance from the 
current position. Crucially, the Hierarchical Interest Encoder's output $\mathbf{z}$ 
serves as a structurally balanced supervision signal to regularize the representational 
geometry of $\mathbf{u}$, and is discarded at inference to preserve the fine-grained 
sequential fidelity of the Sequential Inference Encoder.

\paragraph{Hierarchical Segment Representation}
Given boundary flags, items are assigned to segments via cumulative summation: 
$\mathrm{seg}(i) = \sum_{j \leq i} b_j$, yielding $K$ segments $\{s_1, \dots, s_K\}$. 
Each segment $s_k$ is summarized by a learned additive attention:
\begin{equation}
\mathbf{p}_k = \sum_{i \in s_k} \alpha_i \mathbf{h}_i, \quad
\alpha_i = \frac{\exp(\mathbf{w}^\top \mathbf{h}_i / \sqrt{d})}
               {\sum_{j \in s_k} \exp(\mathbf{w}^\top \mathbf{h}_j / \sqrt{d})},
\end{equation}
where $\mathbf{w} \in \mathbb{R}^d$ is a learned query vector. This pooling step 
deliberately discards intra-segment sequential order, producing a segment matrix 
$\mathbf{P} = [\mathbf{p}_1, \dots, \mathbf{p}_K] \in \mathbb{R}^{K \times d}$ in which 
every segment occupies a structurally equivalent slot. $\mathbf{P}$ is then passed through 
$L_2$ Transformer blocks with MS-RoPE and a causal mask, capturing the temporal 
evolution of intent across segments and producing refined representations 
$\mathbf{P}' \in \mathbb{R}^{K \times d}$.

\paragraph{Segment-Count-Adaptive Multi-Intent Extraction}
\label{sec:interest}
User behavior within a session is rarely driven by a single motive. We extract 
$U_{\max}$ interest-specific summaries from $\mathbf{P}'$ via cross-attention with 
$U_{\max}$ orthogonally-initialized learnable interest queries 
$\mathbf{Q} = [\mathbf{q}_1, \dots, \mathbf{q}_{U_{\max}}] \in \mathbb{R}^{U_{\max} \times d}$:
\begin{equation}
\mathbf{m}_r = \mathrm{CrossAttn}(\mathbf{q}_r,\; \mathbf{P}'), \quad r = 1, \dots, U_{\max}
\end{equation}

Each interest query attends to all $K$ segment 
representations, allowing the aggregated multi-interest vector to integrate evidence from the 
full interaction history and thereby serve as a globally-informed teacher 
signal for the causal Sequential Inference Encoder. Since the number of segments $K$ reflects the diversity of behavioral phases, it 
provides a natural upper bound on the number of distinct user interests. However, 
$K$ tends to overestimate the true number of interest types, as the same intent 
can recur periodically across non-adjacent segments. We therefore gate the number 
of active interest slots by a compression ratio $\rho \in (0, 1]$:
\begin{equation} \label{eq:ueff}
U_{\text{eff}} = \min\!\left(\left\lceil \rho K \right\rceil,\; U_{\max}\right) ,
\end{equation}
where $\rho$ controls the degree of slot compression relative to segment count. 
Setting $\rho = 1$ allocates one slot per segment; smaller values consolidate 
capacity when periodic interest recurrence is expected. Given $U_{\text{eff}}$, 
we select the most activated slots from the $U_{\max}$ candidates rather than 
retaining them by index order.

Intuitively, an interest vector with a larger $\ell_2$ norm has attended 
more selectively to specific segments, whereas a near-zero norm indicates 
that the corresponding query failed to aggregate any coherent behavioral 
signal and should be suppressed. We select the $U_{\text{eff}}$ slots with the largest $\ell_2$ norms, i.e., $\mathcal{A} = \{r : \|\mathbf{m}_r\|_2 \text{ ranks in top-}U_{\text{eff}}\}$.
A compact interest-level user representation is obtained by combining 
element-wise max-pooling and mean-pooling over the active slots, where max-pooling 
captures the most salient interest signal in each dimension across slots and mean-pooling 
retains the overall interest distribution:
\begin{equation}
\mathbf{z} = \operatorname{MaxPool}^{\mathrm{elem}}_{r \in \mathcal{A}}(\mathbf{m}_r) 
           + \frac{1}{U_{\text{eff}}} \sum_{r \in \mathcal{A}} \mathbf{m}_r
\end{equation}
where $[\operatorname{MaxPool}^{\mathrm{elem}}_{r \in \mathcal{A}}(\mathbf{m}_r)]_\ell
= \max_{r \in \mathcal{A}} [\mathbf{m}_r]_\ell$ for each feature dimension $\ell$.

\paragraph{Interest Diversity Regularization}
Although the interest queries $\mathbf{Q}$ are orthogonally initialized, 
soft cross-attention does not prevent the active interest vectors 
$\{\mathbf{m}_r\}_{r=1}^{U_{\text{eff}}}$ from collapsing toward a common 
direction during training, which would cause $\mathbf{z}$ to degenerate into 
a single-interest representation and undermine its role as a structurally 
balanced teacher signal for the contrastive objective. 
To prevent this, we introduce a lightweight diversity regularizer on the 
$\ell_2$-normalized active interest matrix 
$\tilde{\mathbf{M}} \in \mathbb{R}^{U_{\text{eff}} \times d}$:
\begin{equation}
\mathcal{L}_{\text{div}} = 
\left\| \tilde{\mathbf{M}} \tilde{\mathbf{M}}^\top - \mathbf{I} \right\|_F^2 ,
\label{eq:div_loss}
\end{equation}
where $\tilde{\mathbf{m}}_r = \mathbf{m}_r / \|\mathbf{m}_r\|$ and 
$\mathbf{I} \in \mathbb{R}^{U_{\text{eff}} \times U_{\text{eff}}}$ is the 
identity matrix. This loss penalizes pairwise cosine similarity between 
interest vectors, encouraging them to remain approximately orthogonal 
and thereby preserving the multi-interest expressiveness of $\mathbf{z}$.

\subsection{Model Optimization}
\label{sec:training}

We now describe the training objectives. In what follows, a batch contains
$B$ users; subscript $i$ indexes users within the batch. For the $i$-th
user, the Sequential Inference Encoder  produces $\mathbf{u}_i = \mathbf{h}_{N_i}$
and the Hierarchical Interest Encoder produces $\mathbf{z}_i$.

\paragraph{Recommendation Loss}
Let $\mathbf{E}_V \in \mathbb{R}^{|\mathcal{V}| \times d}$ be the item embedding matrix, where each row corresponds to an item vector.
The prediction score distribution over all candidate items is:
\begin{equation}
\hat{\mathbf{y}}_i = \mathrm{Softmax}(\mathbf{E}_V\, \mathbf{u}_i)
\label{eq:predict}
\end{equation}

The recommendation loss is then defined as the full cross-entropy between the predicted distribution and the one-hot ground-truth label:
\begin{equation}
\mathcal{L}_{CE}
= - \sum_{v=1}^{|\mathcal{V}|} \bar{y}_v \log(\hat{y}_v),
\label{eq:pred_loss}
\end{equation}
where $\bar{y}_v = 1$ only for the target item and 0 otherwise, and the sum is over all items in the catalog $\mathcal{V}$. No sampled negatives or in-batch candidate approximation are used.

\paragraph{Contrastive Regularization Loss}
The multi-interest representation $\mathbf{z}_i$ from the Hierarchical Interest Encoder
and the sequential user embedding $\mathbf{u}_i$ from the Sequential Inference Encoder 
constitute two complementary views of the same user state: $\mathbf{u}_i$ captures
fine-grained autoregressive dependencies, while $\mathbf{z}_i$ provides a
structurally balanced summary across all intent phases (see \S\ref{sec:segment}).
We align them with a contrastive loss using in-batch negatives.

Let
$\tilde{\mathbf{u}}_i = \mathbf{u}_i / \|\mathbf{u}_i\|$ and
$\tilde{\mathbf{z}}_i = \mathbf{z}_i / \|\mathbf{z}_i\|$
denote the $\ell_2$-normalized representations of the $i$-th user from
the Sequential Inference Encoder and the Hierarchical Interest Encoder, respectively. The contrastive loss is:
\begin{equation}
\mathcal{L}_{\text{CL}} = -\frac{1}{B} \sum_{i=1}^{B}
\log \frac{
  \exp\!\left(\tilde{\mathbf{u}}_i^\top \tilde{\mathbf{z}}_i \,/\, \eta\right)
}{
  \sum_{j=1}^{B} \exp\!\left(\tilde{\mathbf{u}}_i^\top \tilde{\mathbf{z}}_j \,/\, \eta\right)
},
\label{eq:cl_loss}
\end{equation}
where $\eta$ is a temperature hyperparameter. Each user's sequential
representation $\tilde{\mathbf{u}}_i$ is treated as the anchor, with the
corresponding multi-interest representation $\tilde{\mathbf{z}}_i$ as the
positive and all other $\tilde{\mathbf{z}}_j\ (j \neq i)$ as in-batch negatives.
Gradients from $\mathcal{L}_{\text{CL}}$ flow simultaneously into the Sequential Inference Encoder 
(through $\mathbf{u}_i$) and the Hierarchical Interest Encoder (through $\mathbf{z}_i$), imposing
cross-path geometric regularization without altering the inference-time computation.

\paragraph{Joint Training}
The final training loss combines the recommendation loss, the contrastive
regularizer, and the interest diversity regularizer:
\begin{equation}
\mathcal{L} = \mathcal{L}_{\text{CE}} + \gamma\,\mathcal{L}_{\text{CL}}
            + \beta\,\mathcal{L}_{\text{div}},
\label{eq:total_loss}
\end{equation}
where $\gamma$ controls the strength of the contrastive regularizer and 
$\beta$ controls the diversity penalty. Note that $\mathcal{L}_{\text{div}}$
operates only on the Hierarchical Interest Encoder, whereas
$\mathcal{L}_{\text{CL}}$ aligns the Sequential Inference Encoder and the
Hierarchical Interest Encoder during training. Both auxiliary objectives are
discarded at inference, incurring no additional cost during serving.

\subsection{Complexity Analysis}


\begin{table}[htbp]
\centering
\caption{Summary of Notations}
\label{tab:notation}
\renewcommand{\arraystretch}{1.25}
\begin{tabular}{@{} l p{9cm} @{}}
\toprule
\textbf{Notation} & \textbf{Description} \\
\midrule

$\mathcal{U},\, \mathcal{V}$ & Sets of users and items \\
$\mathcal{S} = \{(v_i, c_i, t_i)\}_{i=1}^{N}$ & Chronologically ordered interaction history of a user \\
$v_i \in \mathcal{V}$ & Item interacted with at position $i$ \\
$c_i$ & Category label of item $v_i$ \\
$t_i$ & Timestamp of the $i$-th interaction \\
$\Delta t_i = t_{i+1} - t_i$ & Inter-event time interval \\
\midrule

$\mathbf{e}(t) \in \mathbb{R}^{d}$ & Latent intent energy vector \\
$\boldsymbol{\delta}_i \in \mathbb{R}^{d}_{+}$ & Dimension-wise decay rate vector at position $i$ \\
$r_i$ & Energy retention ratio \\
$\tau \in (0,1)$ & Boundary threshold \\
$\Lambda_i$ & TPP compensator; integral of $\lambda(t)$ over $(t_i,\, t_{i+1})$ \\
$w(c_i, c_{i+1}) \in (0,1]$ & Soft co-occurrence weight between adjacent categories \\
$\alpha$ & Scaling factor in the soft co-occurrence weight \\
$b_i \in \{0,1\}$ & Binary boundary flag ($b_i=1$ marks the start of a new segment) \\
$K$ & Total number of segments produced by the frozen Soft-TPP \\
\midrule

$U_{\max}$ & Maximum number of interest slots (upper cap on $U_{\text{eff}}$) \\
$\rho \in (0,1]$ & Compression ratio; controls the degree of slot compression relative to segment count \\
$\eta$ & Temperature hyperparameter in the contrastive loss $\mathcal{L}_{\text{CL}}$ \\
$\gamma$ & Weight of the contrastive regularizer in the joint training loss \\
$\beta$ & Weight of the interest diversity regularizer in the joint training loss \\
\bottomrule
\end{tabular}
\end{table}

\begin{algorithm}[t]
\caption{Training Procedure of S2-CAR}
\label{alg:s2car}
\begin{algorithmic}[1]
\renewcommand{\algorithmicrequire}{\textbf{Input:}}
\renewcommand{\algorithmicensure}{\textbf{Output:}}
\REQUIRE Chronologically split prefixes 
$\mathcal{S}_{\mathrm{train}}, \mathcal{S}_{\mathrm{val}}, \mathcal{S}_{\mathrm{test}}$; 
item set $\mathcal{V}$; category set $\mathcal{C}$
\ENSURE Best model parameters $\Theta^*$

\STATE Initialize Soft-TPP parameters $\Theta_{\mathrm{TPP}}=\{\mathbf{E}_C,f_\delta\}$
\STATE Train $\Theta_{\mathrm{TPP}}$ on $\mathcal{S}_{\mathrm{train}}$ by minimizing 
$\mathcal{L}_{\mathrm{TPP}}$ in Eq.~\eqref{eq:tpp_loss}; freeze $\Theta_{\mathrm{TPP}}$

\STATE Initialize recommendation parameters 
$\Theta=\{\Theta_{\mathrm{SIE}},\Theta_{\mathrm{HIE}}\}$ and set 
$\Theta^*\leftarrow\Theta$
\REPEAT
    \FOR{each training prefix $S\in\mathcal{S}_{\mathrm{train}}$}
        \STATE Generate boundary flags $\{b_i\}$ using the frozen Soft-TPP
        \STATE Obtain $\mathbf{u}$ from the SIE and $\mathbf{z}$ from the HIE
        \STATE Update $\Theta$ by minimizing 
        $\mathcal{L}_{\mathrm{CE}}+\gamma\mathcal{L}_{\mathrm{CL}}+\beta\mathcal{L}_{\mathrm{div}}$
    \ENDFOR
    \STATE Update $\Theta^*$ if metrics on $\mathcal{S}_{\mathrm{val}}$ improve
\UNTIL{early stopping is triggered}

\STATE \textbf{return} $\Theta^*$
\end{algorithmic}
\end{algorithm}

Algorithm~\ref{alg:s2car} presents the training procedure for the proposed S2-CAR. The model follows a two-stage design: the Soft-TPP is first pretrained to capture continuous-time intent dynamics and then frozen, after which boundary flags are used to guide both encoders in the main training loop. To avoid split leakage under leave-one-out evaluation, Soft-TPP pretraining uses only the training prefixes. During validation and testing, the frozen Soft-TPP generates boundary flags only from adjacent interactions within the visible prefix of the corresponding split; the held-out target item's identity, timestamp, and category are not used to compute intervals, co-occurrence weights, or boundary flags. In particular, no boundary is generated from the interval between the last visible item and the held-out target. The SIE processes the visible item sequence to produce a fine-grained sequential user embedding, while the HIE constructs a hierarchically balanced multi-interest representation as a contrastive supervision signal.

We analyze the computational complexity of S2-CAR. Given embedding dimension $d$, sequence length $N$, number of segments $K$, maximum interest slots $U_{\max}$, and catalog size $|\mathcal{V}|$, the Soft-TPP incurs $\mathcal{O}(N \cdot d)$ per sequence and is pretrained once and frozen, contributing no cost to the main training loop. The SIE applies $L_1$ Transformer blocks over the full sequence, yielding $\mathcal{O}(N^2 \cdot d)$ (with $L_1$ absorbed as a small constant). MS-RoPE adds linear rotation operations over query and key vectors in each head, which are absorbed into the Transformer constant when the number of heads is fixed. The HIE operates on $K$ segment vectors where $K \ll N$, with a segment-level Transformer costing $\mathcal{O}(K^2 \cdot d)$, cross-attention over $U_{\max}$ interest slots contributing $\mathcal{O}(U_{\max} \cdot K \cdot d)$, and the diversity regularizer adding $\mathcal{O}(U_{\max}^2 \cdot d)$; since both $U_{\max}$ and $L_2$ are small constants, the HIE reduces to $\mathcal{O}(K^2 \cdot d)$. In addition to the encoder cost, the full-catalog prediction layer used for cross-entropy training and full-ranking evaluation requires $\mathcal{O}(|\mathcal{V}| \cdot d)$ scoring per instance, or $\mathcal{O}(B|\mathcal{V}| \cdot d)$ for a mini-batch of size $B$. The joint per-instance complexity is therefore $\mathcal{O}(N^2 \cdot d + K^2 \cdot d + |\mathcal{V}| \cdot d)$ when full softmax scoring is included.

\section{Experiments}

\subsection{Datasets}
\begin{table}[htbp]
\caption{Statistics of the datasets.}
\centering
\resizebox{\textwidth}{!}{
\begin{tabular}{lccccccc}
\toprule[1pt]
\textbf{Dataset} & \textbf{\# Users} & \textbf{\# Items} & \textbf{\# Cats} & \textbf{\# Interacts} & \textbf{Avg. Len} & \textbf{Density} & \textbf{Duration (h)} \\
\midrule
ML-1M  & 6,040  & 3,416   & 18 & 999,611   & 165.50 & 4.84\% & 395.55 \\
Amazon & 65,262 & 106,427 & 28 & 1,540,690 & 23.61 & 0.02\% & 4,128.80 \\
Steam  & 39,795 & 10,587  & 44 & 1,790,393 & 44.99 & 0.42\% & 30,761.08 \\
\bottomrule[1pt]
\end{tabular}%
}
\label{tab:dataset_stats}
\end{table}

We evaluate our method on three publicly available benchmarks that span diverse application domains and exhibit varying levels of interaction sparsity, ensuring a comprehensive and rigorous assessment.

\begin{itemize}
    \item \textbf{ML-1M}~\citep{harper2015movielens} is a widely adopted movie rating dataset from the MovieLens platform, comprising one million explicit ratings assigned by approximately 6,000 users to around 4,000 films. Following common sequential recommendation practice, ratings are treated as implicit interactions and ordered by timestamp. Its relatively dense interaction structure makes it a standard testbed for sequential recommendation research.

    \item \textbf{Amazon}~\citep{he2016ups} is a large-scale product review collection spanning multiple merchandise categories on the Amazon e-commerce platform. To balance data scale with computational tractability, we retain the most recent 2~million interactions aggregated across all categories before preprocessing, preserving both temporal ordering and cross-category diversity.

    \item \textbf{Steam}~\citep{kang2018self} is a video game interaction dataset collected from the Steam gaming platform, capturing user purchase and play histories. Its domain-specific nature and moderate sparsity provide a complementary evaluation setting to the other two benchmarks.
\end{itemize}

All datasets are preprocessed following the filtering procedure that retains users with at least 20 interactions and items with at least 5 interactions. Category metadata is mapped before model training and then kept fixed across all splits. Data partitioning adheres to the widely adopted leave-one-out protocol~\citep{kang2018self,li2021hyperbolic}: the most recent interaction of each user is reserved for testing, the second-most-recent for validation, and all remaining interactions constitute the training set. This chronological split strictly respects the temporal ordering of interactions to prevent data leakage. For each user, the validation instance uses only the training prefix to predict the validation item, while the test instance uses the training plus validation prefix to predict the test item. Soft-TPP pretraining is performed only on training prefixes. When generating boundary flags for validation and testing, the frozen detector is applied only to adjacent events inside the visible prefix, so the held-out target's item identity, timestamp, and category are not used to compute $\Delta t_i$, $w(c_i,c_{i+1})$, or boundary flags. Thus, the boundary sequence available to S2-CAR at evaluation time is causally determined by observed interactions only. Summary statistics for all three datasets are reported in~\Cref{tab:dataset_stats}. Notably, the three datasets exhibit substantially different temporal scales: ML-1M has a compact average sequence duration of 395.55 hours, Amazon spans 4,128.80 hours on average, and Steam extends to 30,761.08 hours, reflecting the long-term nature of gaming engagement.

\subsection{Baselines}

We compare our proposed method against a representative and diverse set of sequential recommendation models, ranging from classical approaches to state-of-the-art contrastive learning frameworks. We compare against 13 baselines spanning three categories: 
general sequential recommenders (GRU4Rec, Caser, SASRec, BERT4Rec, DSSRec), 
augmentation and multi-intent methods (MiaSRec, CL4SRec, DuoRec, ICSRec, BASRec), 
and temporal-aware recommenders (TiSASRec, DT-GAT, TLSRec). We also summarize the features of different baselines and our proposed model in Table~\ref{tab:feature_compare}.
\begin{itemize}
    \item \textbf{GRU4Rec}~\citep{hidasi2015session} is among the earliest deep sequential recommenders, employing gated recurrent units to encode the temporal dynamics of user sessions. It introduces session-parallel mini-batching and ranking-based loss functions tailored to the recommendation setting.

    \item \textbf{Caser}~\citep{tang2018personalized} frames sequential recommendation as an image recognition problem by treating the embedding matrix of recent items as an ``image'' and applying both horizontal and vertical convolutional filters to capture point-level and union-level transition patterns at multiple granularities.

    \item \textbf{DSSRec}~\citep{ma2020disentangled} proposes a sequence-to-sequence training strategy based on latent self-supervision and disentanglement. Rather than predicting only the next item, it reconstructs representations of future subsequences in the latent space, and constructs training pairs from subsequences that share the same underlying user intent, enabling the model to capture longer-term and more diverse user preferences.

    \item \textbf{SASRec}~\citep{kang2018self} introduces an unidirectional self-attention architecture that adaptively attends to relevant items in a user's history to predict the next interaction. Its scalability and strong empirical performance have established it as a canonical sequential recommendation baseline.

    \item \textbf{BERT4Rec}~\citep{sun2019bert4rec} adapts the bidirectional encoder representations from the BERT language model to sequential recommendation via a cloze-style masked item prediction task, enabling the model to leverage both left- and right-context signals when learning item representations.


    \item \textbf{CL4SRec}~\citep{xie2022contrastive} is a contrastive framework for sequential recommendation that introduces three sequence-level data augmentation operators, random cropping, item masking, and subsequence reordering, and enforces representation consistency between augmented views through an InfoNCE-style contrastive objective.

    \item \textbf{DuoRec}~\citep{qiu2022contrastive} addresses the representation degeneration problem in transformer-based sequential models by combining a model-level unsupervised augmentation based on dropout-induced stochastic encoders with a supervised contrastive objective that pulls together sequences sharing the same target item.


    \item \textbf{MiaSRec}~\citep{choi2024multi} augments the standard transformer architecture with a multi-intent module that routes each item to several intent embeddings, allowing the model to disentangle and concurrently track multiple evolving user intents within a single sequence encoder.

    \item \textbf{ICSRec}~\citep{qin2024intent} decomposes user interaction histories into intent-homogeneous subsequences by clustering items into latent intent categories, and then applies contrastive learning at the intent level to explicitly align representations of sequences sharing the same underlying user purpose.

    \item \textbf{BASRec}~\citep{dang2025augmenting} augments sequential recommendation with balanced relevance and diversity, constructing augmented training signals that preserve target relevance while improving behavioral diversity.

    \item \textbf{TiSASRec}~\citep{li2020time} {\color{black} employs a time-aware self-attention mechanism that incorporates both absolute positional information and relative time intervals between user interactions, enabling the model to capture temporal dynamics within the sequence.}

    \item \textbf{TLSRec}~\citep{chen2022time} segments interaction histories into sessions and applies a hierarchical self-attention network to capture preferences at both the session and intra-session levels, fusing them through a neural time gate that adaptively weights long- and short-term signals based on the elapsed time since the user's last interaction.

    \item \textbf{DT-GAT}~\citep{guo2026dual} proposes a dual-channel temporal graph attention architecture that jointly models item-level and sequence-level dynamics, combined with a multi-temporal window mechanism to produce user representations.

\end{itemize}

\subsection{Evaluation Metrics}
We evaluate all methods using two standard top-$K$ ranking metrics widely
adopted in sequential recommendation: Recall (R@$K$) and Mean Reciprocal
Rank (MRR@$K$), with $K \in \{5, 10\}$.
R@$K$ measures retrieval effectiveness by computing the fraction of test
cases in which the ground-truth item appears within the top-$K$ ranked
candidates:
\begin{equation}
    \text{R@}K = \frac{1}{|\mathcal{U}|} \sum_{i=1}^{|\mathcal{U}|}
    \mathbb{I}(r_i \leq K),
\end{equation}
where $\mathcal{U}$ denotes the set of test users, $r_i$ is the rank
assigned to the ground-truth item for the $i$-th user, and $\mathbb{I}(\cdot)$
is the indicator function returning 1 if the condition holds and 0 otherwise.
R@$K$ treats all positions within the top-$K$ list equally and thus
reflects coverage without regard to ranking position.
MRR@$K$ complements R@$K$ by incorporating positional sensitivity,
rewarding methods that place the ground-truth item closer to the top of
the ranked list:
\begin{equation}
    \text{MRR@}K = \frac{1}{|\mathcal{U}|} \sum_{i=1}^{|\mathcal{U}|}
    \frac{\mathbb{I}(r_i \leq K)}{r_i}.
\end{equation}
Since each test case contains exactly one ground-truth item, the reciprocal
rank $1/r_i$ is non-zero only when the ground-truth item falls within the
top-$K$ list, making the above expression equivalent to the standard
truncated form. Higher MRR@$K$ values indicate that the ground-truth item
not only appears in the top-$K$ list but is consistently ranked near the
top, thereby reflecting the quality of the ranking more faithfully than
R@$K$ alone.

\subsection{Experiment Settings}

All models are implemented in PyTorch and trained on a single NVIDIA RTX~A5000 GPU (24~GB VRAM). To ensure fair and reproducible comparisons, we enforce a unified set of core hyperparameters across all baselines and our proposed method: a hidden embedding dimensionality of 100, a mini-batch size of 100, and $\ell_2$-normalized parameter initialization to stabilize early training dynamics. All models are optimized with the Adam optimizer at a fixed learning rate of $10^{-3}$, and training is terminated via early stopping with a patience of 20~epochs monitored on validation R@10. All baseline numbers are reproduced under this unified protocol, rather than taken from original papers, to ensure a fair comparison under the same evaluation conditions.

To eliminate the confounding effect of different negative-sampling strategies and sampled candidate sizes in sequential recommendation evaluation, we use full cross-entropy over the entire item set $\mathcal{V}$, rather than sampled softmax or negative sampling. During evaluation, each test instance is ranked against the complete candidate set of size $|\mathcal{V}|$ (i.e., all items in the catalog), and no sampled negatives are used for fair comparison~\citep{krichene2020sampled}. 

For S2-CAR, additional hyperparameters are set as follows. The Soft-TPP pretraining stage uses the same Adam optimizer, learning rate, batch size, and 5-run seed protocol as the main model unless otherwise specified, and is run for 5 epochs before freezing. The TPP scaling factor $\alpha$ is fixed at 10. The contrastive loss weight $\gamma$ is searched over ${0.001, 0.01, 0.1, 0.5}$, the diversity weight $\beta$ over ${0.001, 0.005, 0.01, 0.2}$ and selected as 0.005 according to validation R@10, the compression ratio $\rho$ over ${0.25, 0.5, 0.75, 1.0}$, and the energy threshold $\tau$ over ${0.5, 0.7, 0.8, 0.9}$. The contrastive temperature is fixed at 0.5. The SIE and HIE each use 2 Transformer layers and 10 attention heads. The maximum interest slot count $U_{\max}$ is set to 50. For MS-RoPE, per-head frequency scales are initialized on a log-uniform grid spanning $[1.0, 10.0]$ across the 10 heads, allowing each head to specialize in positional dependencies at a different granularity.

To account for stochastic variability in model training, all experiments are repeated independently 5~times with different random seeds. We report the mean performance across runs and provide 95\% confidence intervals for the best and second-best methods on each dataset to indicate the variability of observed improvements.

To better evaluate the performance of the proposed S2-CAR, we design experiments to answer the following research questions:

\begin{itemize}[leftmargin=*]
    \item \textbf{RQ1 (Overall Performance):} 
    Does S2-CAR consistently outperform state-of-the-art 
    sequential recommenders across diverse benchmarks?

    \item \textbf{RQ2 (Plug-in Generalizability):} 
    Is the proposed energy-based TPP segmentation a broadly 
    applicable design? Does integrating it into existing 
    sequential models as a plug-in module yield consistent 
    performance gains?

    \item \textbf{RQ3 (Ablation Analysis):} 
    How do the individual components of S2-CAR, the 
    Context-Aware Soft-TPP segmentation, the 
    Segment-Count-Adaptive Multi-Intent Extraction, and the 
    Hierarchical Interest Encoder, each contribute to 
    overall recommendation performance?

    \item \textbf{RQ4 (Hyperparameter Sensitivity):} 
    How sensitive is S2-CAR to key hyperparameters, 
    including the energy threshold $\tau$, the compression 
    ratio $\rho$, and the loss weights $\gamma$?

    \item \textbf{RQ5 (Computational Efficiency):} 
    Is the additional computational overhead introduced 
    by the two-stage training procedure of S2-CAR 
    acceptable for practical deployment?

    \item \textbf{RQ6 (Middle Information Utilization):} 
    {\color{black}Can S2-CAR effectively utilize intent-relevant signals embedded in the middle of interaction sequences?}
\end{itemize}

\subsection{Overall Performance(RQ1)}\label{sec:exp1}

\begin{table*}[htbp]
{\color{black}
\centering
\caption{Feature comparison across sequential recommendation models. A checkmark (\ding{51}) indicates the model incorporates the corresponding technique, while a cross mark (\ding{55}) indicates it does not. Feature labels are assigned according to the definitions in this caption; adaptive intent segmentation denotes learned or data-adaptive boundary inference rather than fixed window or heuristic session partitioning.}
\label{tab:feature_compare}
\resizebox{\textwidth}{!}{
\begin{tabular}{l|ccc|cc}
\toprule[1pt]
\multirow{2}{*}{\textbf{Models}} 
  & \multicolumn{3}{c|}{\textbf{Intent Modeling}} 
  & \multicolumn{2}{c}{\textbf{Modeling Techniques}} \\
\cmidrule(lr){2-4} \cmidrule(lr){5-6}
& \textbf{Temporal Dynamics}
& \textbf{Adaptive Intent Segmentation}
& \textbf{Multi-Interest Modeling}
& \textbf{Hierarchical Encoding}
& \textbf{Category-Aware Modeling} \\
\midrule
\textbf{GRU4Rec}
& \ding{55} & \ding{55} & \ding{55} & \ding{55} & \ding{55} \\
\textbf{Caser}
& \ding{55} & \ding{55} & \ding{55} & \ding{55} & \ding{55} \\
\textbf{BERT4Rec}
& \ding{55} & \ding{55} & \ding{55} & \ding{55} & \ding{55} \\
\textbf{SASRec}
& \ding{55} & \ding{55} & \ding{55} & \ding{55} & \ding{55} \\
\textbf{DSSRec}
& \ding{51} & \ding{55} & \ding{55} & \ding{55} & \ding{55} \\
\midrule
\textbf{CL4SRec}
& \ding{55} & \ding{55} & \ding{55} & \ding{55} & \ding{55} \\

\textbf{DuoRec}
& \ding{55} & \ding{55} & \ding{55} & \ding{55} & \ding{55} \\
\textbf{MiaSRec}
& \ding{55} & \ding{55} & \ding{51} & \ding{55} & \ding{55} \\
\textbf{ICSRec}
& \ding{55} & \ding{55} & \ding{51} & \ding{55} & \ding{55} \\
\textbf{BASRec}
& \ding{55} & \ding{55} & \ding{55} & \ding{55} & \ding{55} \\
\midrule
\textbf{TiSASRec}
& \ding{51} & \ding{55} & \ding{55} & \ding{55} & \ding{55} \\
\textbf{TLSRec}
& \ding{51} & \ding{55} & \ding{55} & \ding{51} & \ding{55} \\
\textbf{DT-GAT}
& \ding{51} & \ding{55} & \ding{51} & \ding{55} & \ding{55} \\
\midrule
\textbf{S2-CAR}
& \ding{51} & \ding{51} & \ding{51} & \ding{51} & \ding{51} \\
\bottomrule[1pt]
\end{tabular}
}
}
\end{table*}

\begin{table*}[htbp]
\centering
\setlength{\tabcolsep}{2pt}
\caption{Performance comparison on three datasets. We report 95\% confidence intervals (5 runs) for second-best and best methods. Best results are \textbf{bold}, and second-best are \underline{underlined}.}
\label{tab:main_result}
\resizebox{\textwidth}{!}{%
\begin{tabular}{l|cccc|cccc|cccc}
\toprule
\multirow{2.5}{*}{\textbf{Method}}
  & \multicolumn{4}{c|}{\textbf{ML-1M}}
  & \multicolumn{4}{c|}{\textbf{Amazon}}
  & \multicolumn{4}{c}{\textbf{Steam}} \\
  \cmidrule(lr){2-5} \cmidrule(lr){6-9} \cmidrule(lr){10-13}
& \textbf{R@5} & \textbf{R@10} & \textbf{MRR@5} & \textbf{MRR@10}
& \textbf{R@5} & \textbf{R@10} & \textbf{MRR@5} & \textbf{MRR@10}
& \textbf{R@5} & \textbf{R@10} & \textbf{MRR@5} & \textbf{MRR@10} \\
\midrule
GRU4Rec & 2.44 & 3.29 & 1.44 & 1.56 & 1.95 & 2.85 & 1.15 & 1.35 & 7.48 & 10.23 & 5.19 & 5.50 \\
Caser   & 6.89 & 11.27 & 3.59 & 4.55 & 2.29 & 3.17 & 1.48 & 1.59 & 8.89 & 11.74 & 6.64 & 7.01 \\

\midrule
DSSRec  & 7.21 & 14.35 & 3.11 & 3.97 & 2.23 & 3.68 & 0.62 & 0.72 & 8.20 & 12.50 & 6.88 & 7.39 \\
BERT4Rec & 9.96 & 18.55 & 4.18 & 5.33 & 1.20 & 2.27 & 0.51 & 0.64 & 9.58 & 12.74 & 7.08 & 7.49 \\
SASRec   & 10.19 & 16.63 & 4.92 & 5.75 & 2.76 & 4.23 & 1.41 & 1.61 & 9.82 & 13.05 & 7.13 & 7.59 \\

\midrule
CL4SRec & 9.95 & 16.30 & 5.08 & 5.92 & 2.89 & 4.15 & 1.62 & 1.84 & 9.37 & 12.71 & 6.74 & 7.17 \\
DuoRec  & 11.58 & 17.41 & 5.71 & 5.94 & 1.51 & 2.52 & 0.77 & 0.91 & 7.31 & 10.66 & 4.55 & 5.00 \\
MiaSRec  & 11.80 & 19.75 & 5.50 & 6.50 & 2.20 & 3.41 & 1.15 & 1.31 & 9.71 & 12.81 & 7.16 & 7.57 \\
ICSRec  & 10.70 & 17.25 & 5.33 & 6.19 & 2.80 & 4.32 & 1.47 & 1.67 & 9.70 & 13.28 & 7.12 & 7.60 \\
BASRec  & \underline{15.75$\pm$0.14} & \underline{24.75$\pm$0.19} & 8.26 & 9.44 & 2.75 & 4.27 & 1.44 & 1.64 & \underline{9.93$\pm$0.11} & \underline{13.41$\pm$0.13} & \underline{7.21$\pm$0.09} & \underline{7.64$\pm$0.07} \\
\midrule

TiSASRec & 14.09 & 22.10 & 7.12 & 8.15 & 2.48 & 3.91 & 1.28 & 1.46 & 8.96 & 12.52 & 6.41 & 6.87 \\

TLSRec  & 15.50 & 24.41 & \underline{9.18$\pm$0.09} & \underline{9.94$\pm$0.11} & \underline{3.38$\pm$0.11} & \underline{5.18$\pm$0.06} & \underline{1.73$\pm$0.05} & \underline{1.97$\pm$0.09} & 9.75 & 13.30 & 6.80 & 7.28 \\
DT-GAT  & 9.88 & 15.38 & 4.75 & 5.71 & 2.71 & 4.92 & 1.44 & 1.67 & 8.26 & 11.64 & 5.96 & 6.42 \\
\midrule
\textbf{S2-CAR}
  & \textbf{20.03$\pm$0.16} & \textbf{28.74$\pm$0.20} & \textbf{11.92$\pm$0.12} & \textbf{13.07$\pm$0.14}
  & \textbf{4.04$\pm$0.05} & \textbf{5.74$\pm$0.07} & \textbf{2.30$\pm$0.03} & \textbf{2.51$\pm$0.05}
  & \textbf{10.84$\pm$0.10} & \textbf{14.41$\pm$0.14} & \textbf{7.86$\pm$0.08} & \textbf{8.32$\pm$0.09} \\
\midrule
\textbf{Improv.}
  & +27.17\% & +16.12\% & +29.85\% & +31.49\%
  & +19.53\% & +10.81\% & +32.95\% & +27.41\%
  & +9.16\%  & +7.46\%  & +9.02\%  & +8.90\% \\
\bottomrule
\end{tabular}%
}
\end{table*}

Table~\ref{tab:main_result} reports R@5, R@10, MRR@5, and MRR@10 for all
methods across the ML-1M, Amazon, and Steam datasets.
S2-CAR achieves the best mean scores on every metric and dataset. The reported
confidence intervals for the best and second-best methods indicate that the
observed gains are stable across repeated runs.

GRU4Rec and Caser represent an earlier generation of sequential recommenders.
GRU4Rec encodes interaction histories through gated recurrent units but treats
each time step uniformly, which limits its sensitivity to the varying importance
of individual interactions.
Caser instead applies convolutional filters over a fixed-length item window to
detect local transition patterns; because the receptive field is bounded, it
struggles to propagate long-range dependencies.
Both models trail every attention-based method by a wide margin, confirming that
simple recurrence and local convolution are insufficient for capturing the
heterogeneous dynamics present in real-world interaction logs.

Among the augmentation and intent-aware methods, CL4SRec, DuoRec, ICSRec, and BASRec each exhibit dataset-specific strengths: BASRec is the strongest method in this group on ML-1M and Steam, yet none transfers uniformly across all three benchmarks.

Among the temporal and multi-interest methods, TiSASRec outperforms SASRec, confirming that explicit temporal modeling benefits dense, long-horizon logs, but its gains do not transfer to sparser datasets where irregular time-gaps reduce signal reliability. TLSRec's second best performance on Amazon suggests that fixed temporal window segmentation helps when behavioral phases are temporally well-separated. However, its fixed gap-based criterion limits generalization to datasets where intent transitions are less aligned with raw time-gaps. Moreover, MiaSRec's fixed-capacity interest slots struggle to disambiguate the increasingly complex interaction patterns in long histories, while DT-GAT's graph-based aggregation suffers from over-smoothing as sequence length grows, causing distinct behavioral signals to collapse into indistinguishable representations.
S2-CAR maintains a consistent advantage in both retrieval coverage and ranking quality throughout. Notably, the largest relative MRR@5 gain appears on Amazon, while ML-1M has the longest average sequence length (165.50). This suggests that the improvement is not explained by sequence length alone, but also by sparsity, temporal scale, and category heterogeneity. The more modest but still consistent improvements on Steam (avg. length 44.99) further indicate that the approach generalizes beyond the densest long-sequence setting. All results are reported as the mean across 5 independent runs; the reported confidence intervals for the best and second-best methods indicate stable margins across runs.

\subsection{Plug-in Generalizability (RQ2)}\label{sec:exp3}
To evaluate whether the proposed Soft-TPP segmentation module
is broadly applicable beyond the full S2-CAR framework, we integrate it as a
plug-in component into three representative sequential recommendation backbones:
CoSeRec~\citep{liu2021contrastive} (augmentation-based), SURGE~\citep{chang2021sequential} (graph-based), and
DiffuRec~\citep{li2023diffurec} (diffusion-based). For each backbone, the proposed energy-based boundary is incorporated as a learnable embedding added to the input item embeddings, while all other architectural components and hyperparameters remain unchanged.
Results are reported in~\Cref{tab:backbone}.

Across all three backbones and all three datasets, integrating the TPP
segmentation module yields consistent improvements,
confirming that the  Soft-TPP provides a broadly useful
structural signal rather than one that is tailored exclusively to the S2-CAR
architecture.

The gains are pronounced on the Amazon dataset; we attribute this to the sparser and more heterogeneous nature of the Amazon interaction logs: users on e-commerce platforms tend to exhibit sharper and more frequent intent transitions across product categories, making accurate boundary detection particularly valuable. By contrast, gains on ML-1M are more modest (e.g., $2.76\%$ R@10 for both CoSeRec and DiffuRec), consistent with the observation that the denser and more temporally compact MovieLens sequences leave less room for segmentation
to disambiguate overlapping intents. Among the three backbones, SURGE benefits most from the plug-in module. SURGE originally
constructs an item-item interest graph based on pairwise co-occurrence; without
explicit intent boundaries, cross-phase co-occurrences may introduce spurious edges
that conflate distinct behavioral phases. The injected TPP boundary embedding helps
SURGE distinguish cross-boundary co-occurrences through boundary-aware item
representations, thereby reducing the impact of such cross-phase edges without
changing the graph construction procedure. CoSeRec and DiffuRec exhibit smaller but consistent gains,
indicating that augmentation and diffusion-based methods
also benefit from the design.

\begin{table}[htbp]
  \centering
  \caption{Plug-and-Play comparison across three datasets. Integrating our TPP segmentation module into existing sequential backbones yields consistent improvements.}
  \label{tab:backbone}
  \resizebox{0.8\textwidth}{!}{%
  \begin{tabular}{l|cccccc}
    \toprule
    \multirow{2.5}{*}{\textbf{Method}} 
      & \multicolumn{2}{c}{\textbf{ML-1M}} 
      & \multicolumn{2}{c}{\textbf{Amazon}} 
      & \multicolumn{2}{c}{\textbf{Steam}} \\
    \cmidrule(lr){2-3} \cmidrule(lr){4-5} \cmidrule(lr){6-7}
      \multicolumn{1}{c|}{}
    & \textbf{R@10} & \textbf{MRR@10}
    & \textbf{R@10} & \textbf{MRR@10}
    & \textbf{R@10} & \textbf{MRR@10} \\
    \midrule
    CoSeRec                  & 20.99 & 7.84  & 4.23  & 1.75  & 12.88 & 7.65  \\
    \quad \textbf{+ Ours}    & 21.57 & 8.11  & 4.69  & 1.92  & 13.19 & 7.84  \\
    \quad Improv.            & +2.76\% & +3.44\% & +10.87\% & +9.71\% & +2.41\% & +2.48\% \\
    \midrule
    SURGE                    & 16.59 & 6.21  & 3.81  & 1.94  & 11.92 & 6.19  \\
    \quad \textbf{+ Ours}    & 17.72 & 6.93  & 4.24  & 2.30  & 12.31 & 6.57  \\
    \quad Improv.            & +6.81\% & +11.59\% & +11.29\% & +18.56\% & +3.27\% & +6.14\% \\
    \midrule
    DiffuRec                 & 24.31 & 9.19  & 4.35  & 1.78  & 13.36 & 7.19  \\
    \quad \textbf{+ Ours}    & 24.98 & 9.47  & 5.01  & 1.95  & 13.61 & 7.45  \\
    \quad Improv.            & +2.76\% & +3.05\% & +15.17\% & +9.55\% & +1.87\% & +3.62\% \\
    \bottomrule
  \end{tabular}%
  }
\end{table}

\subsection{Ablation Analysis (RQ3)}
\label{sec:ablation}

To examine how each component contributes to S2-CAR, we conduct an ablation study on all three datasets. The variants in \Cref{tab:ablation_s2car} modify one architectural factor at a time while keeping the remaining training and evaluation settings unchanged.

We first examine the two encoder paths. Removing the Hierarchical Interest Encoder (\textbf{w/o HIE}) consistently reduces performance, indicating that the segment-level path provides useful cross-segment supervision beyond the causal sequential representation. Using the Hierarchical Interest Encoder alone for inference (\textbf{w/o SIE}) leads to a larger decline, since segment pooling weakens intra-segment order information and short-range transition signals. These results support the use of the Sequential Inference Encoder for fine-grained next-item prediction and the Hierarchical Interest Encoder as a structurally balanced training signal.

We then evaluate the boundary construction mechanism, with particular attention to whether the observed gains mainly come from access to category and temporal side information. Inspired by fixed temporal segmentation \citep{chen2022time}, we replace Soft-TPP boundaries with fixed 48-hour segmentation (\textbf{TPP $\rightarrow$ Fixed}), which leads to consistent performance degradation, suggesting that a global time-gap rule is insufficient for capturing user-specific intent shifts. To further separate the effect of side information from the effect of Soft-TPP boundary modeling, we introduce a metadata-only variant (\textbf{w/o TPP}). In this variant, the Soft-TPP module is removed, while temporal embeddings~\citep{guo2026dual} and category embeddings are still provided to the model. Specifically, we add these two side information embeddings to the item embedding at each position, so that the model retains access to temporal and category signals without using TPP-based boundary generation. Thus, \textbf{w/o TPP} does not use continuous-time energy decay, the TPP likelihood, or the energy-retention boundary criterion. Its performance is significantly below the full model. This pattern suggests that category and temporal information are useful, but they do not fully explain the gains of S2-CAR. The comparison provides a controlled test showing that the Soft-TPP boundary signal contributes beyond both heuristic time segmentation and direct temporal-category metadata injection.

Replacing Multi-Scale Rotary Position Embedding with standard absolute position encoding (\textbf{w/o MS-RoPE}) also leads to lower performance, showing that multi-scale relative positional information is useful when user histories contain both short-range transitions and longer-range preference recurrence. Removing the diversity regularizer (\textbf{w/o $\mathcal{L}_{\text{div}}$}) causes a smaller but consistent decline, which is consistent with the role of this term in discouraging collapse among active interest vectors.

We next compare different ways of using the Hierarchical Interest Encoder output. Directly summing $\mathbf{u}_i$ and $\mathbf{z}_i$ before scoring (\textbf{$\mathcal{L}_{\text{CL}} \rightarrow$ Sum}) performs worse than contrastive alignment, suggesting that the hierarchical representation is better used as a training-time regularizer than as an additional inference feature. Replacing the contrastive objective with a gated fusion scheme (\textbf{$\mathcal{L}_{\text{CL}} \rightarrow$ Gated}) shows a similar pattern. These results indicate that forcing the two representations into the same scoring vector may introduce noise, whereas contrastive alignment preserves the causal inference path while still transferring segment-level structure during training.

Finally, we assess the segment summarization and slot allocation designs. Replacing learned segment attention with mean pooling (\textbf{SegAttn $\rightarrow$ Mean}) yields a moderate drop, implying that items within the same segment do not contribute equally to the segment intent. Fixing the number of active interest slots to five (\textbf{DSlot $\rightarrow$ Fixed}) also reduces performance, which suggests that adapting the number of active slots to the inferred segment count better reflects variation in behavioral complexity across users.

\begin{table*}[htbp]
\centering
\caption{Ablation study results.}
\label{tab:ablation_s2car}
\resizebox{0.82\textwidth}{!}{%
\begin{tabular}{l|cc|cc|cc}
\toprule[1pt]
\multicolumn{1}{c|}{\multirow{2.5}{*}{\textbf{Variant}}}
    & \multicolumn{2}{c|}{\textbf{ML-1M}}
    & \multicolumn{2}{c|}{\textbf{Amazon}}
    & \multicolumn{2}{c}{\textbf{Steam}} \\
\cmidrule(lr){2-3} \cmidrule(lr){4-5} \cmidrule(lr){6-7}
\multicolumn{1}{c|}{}
    & \textbf{R@10} & \textbf{MRR@10}
    & \textbf{R@10} & \textbf{MRR@10}
    & \textbf{R@10} & \textbf{MRR@10} \\
\midrule

{w/o HIE}
    & 26.38 & 11.27 & 4.82 & 2.03 & 13.13 & 7.65 \\
{w/o SIE}
    & 21.34 & 9.12  & 4.13 & 1.74 & 11.47 & 6.41 \\

\midrule
{w/o MS-RoPE}
    & 27.13 & 12.21 & 5.41 & 2.29 & 13.74 & 7.91 \\
{w/o $\mathcal{L}_{\text{div}}$}
    & 28.32 & 12.69 & 5.48 & 2.37 & 14.06 & 8.14 \\
{TPP $\rightarrow$ Fixed}
    & 26.91 & 12.24 & 5.26 & 2.18 & 13.83 & 7.94 \\
{w/o TPP}
    & 26.57 & 11.96 & 5.08 & 2.09 & 13.61 & 7.79 \\

\midrule
$\mathcal{L}_{\text{CL}}$ $\rightarrow$ Sum
    & 26.14 & 11.18 & 5.07 & 2.14 & 13.21 & 7.51 \\
$\mathcal{L}_{\text{CL}}$ $\rightarrow$ Gated
    & 26.67 & 11.73 & 5.26 & 2.24 & 13.54 & 7.78 \\

\midrule
SegAttn $\rightarrow$ Mean
    & 27.83 & 12.71 & 5.63 & 2.43 & 14.17 & 8.22 \\
DSlot $\rightarrow$ Fixed
    & 27.19 & 12.29 & 5.51 & 2.36 & 13.89 & 8.07 \\

\midrule
\textbf{S2-CAR}
    & \textbf{28.74} & \textbf{13.07}
    & \textbf{5.74}  & \textbf{2.51}
    & \textbf{14.41} & \textbf{8.32} \\
\bottomrule[1pt]
\end{tabular}%
}
\end{table*}

\subsection{Hyperparameter Sensitivity (RQ4)}\label{sec:exp5}

\begin{figure*}[htbp]
    \centering
    \includegraphics[width=0.25\linewidth]{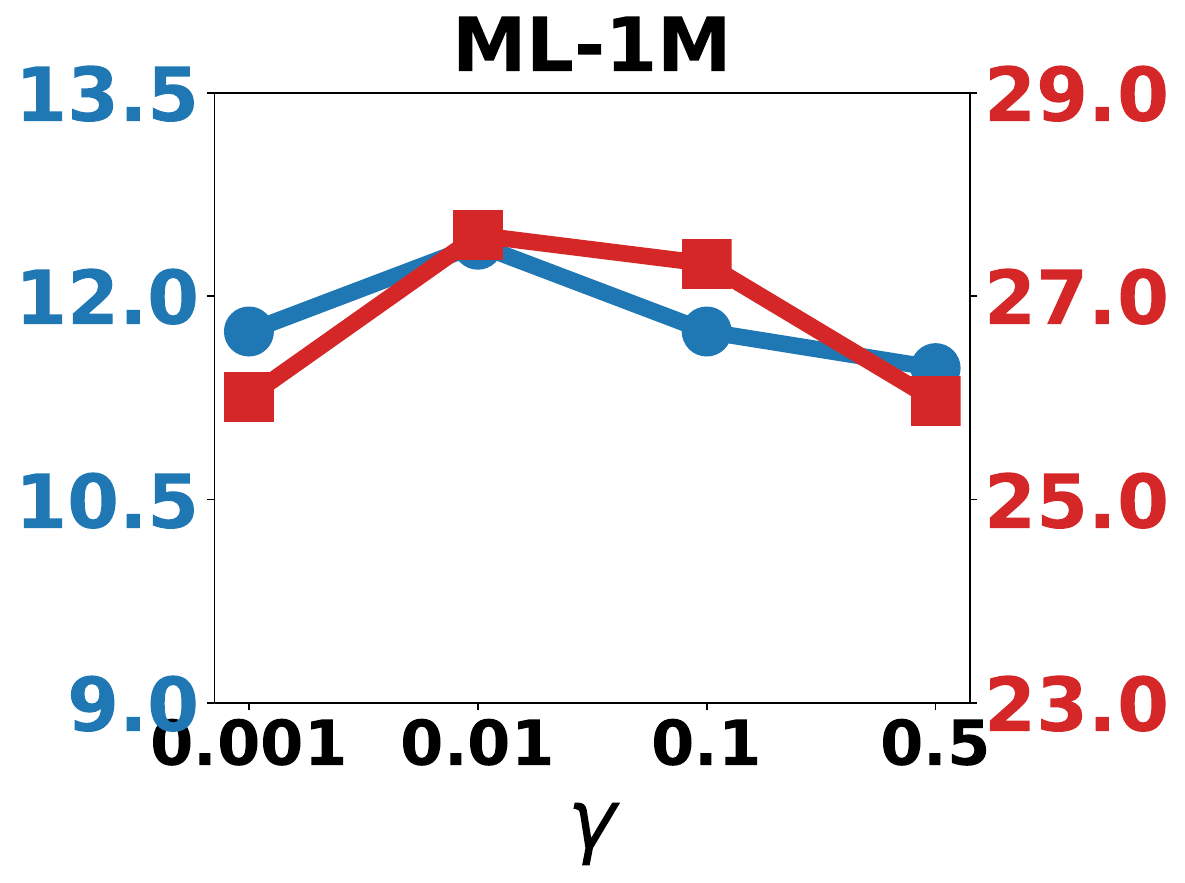}
    \includegraphics[width=0.25\linewidth]{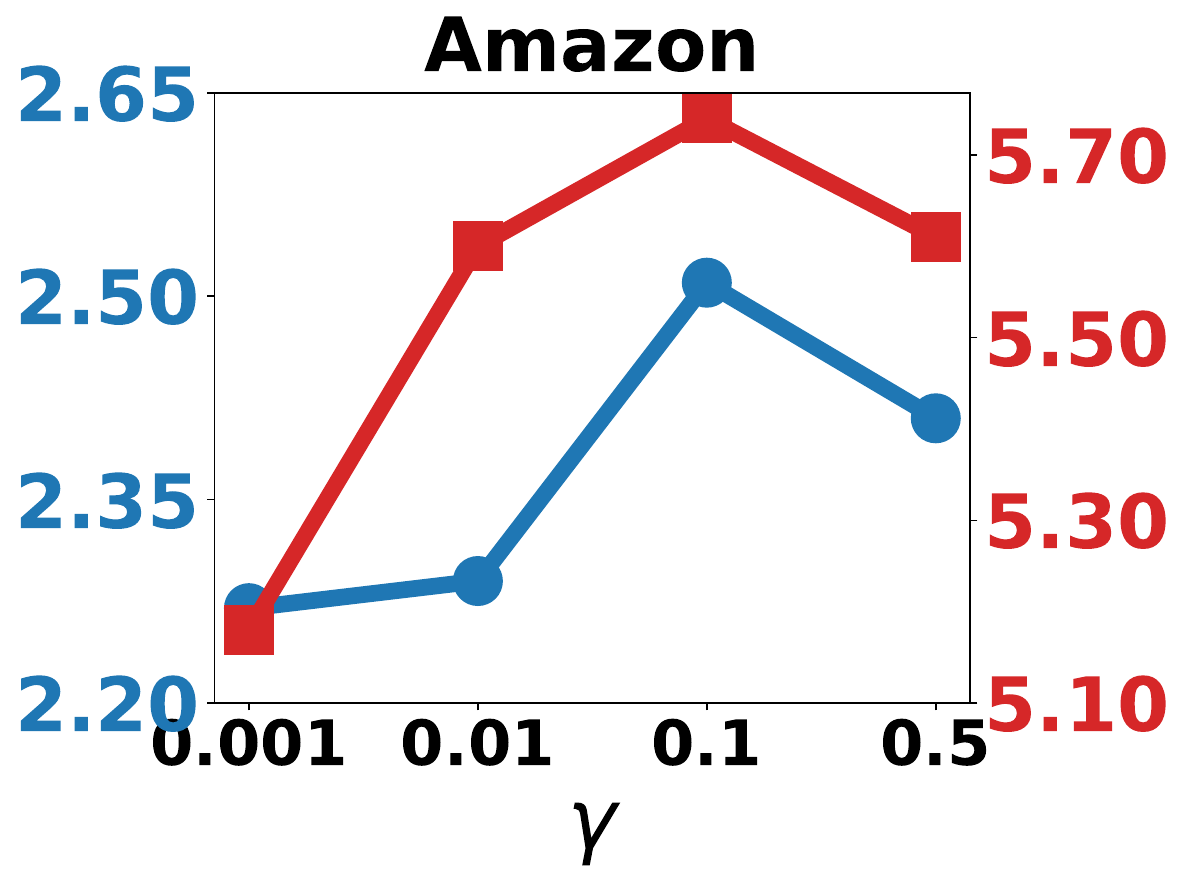}
    \includegraphics[width=0.25\linewidth]{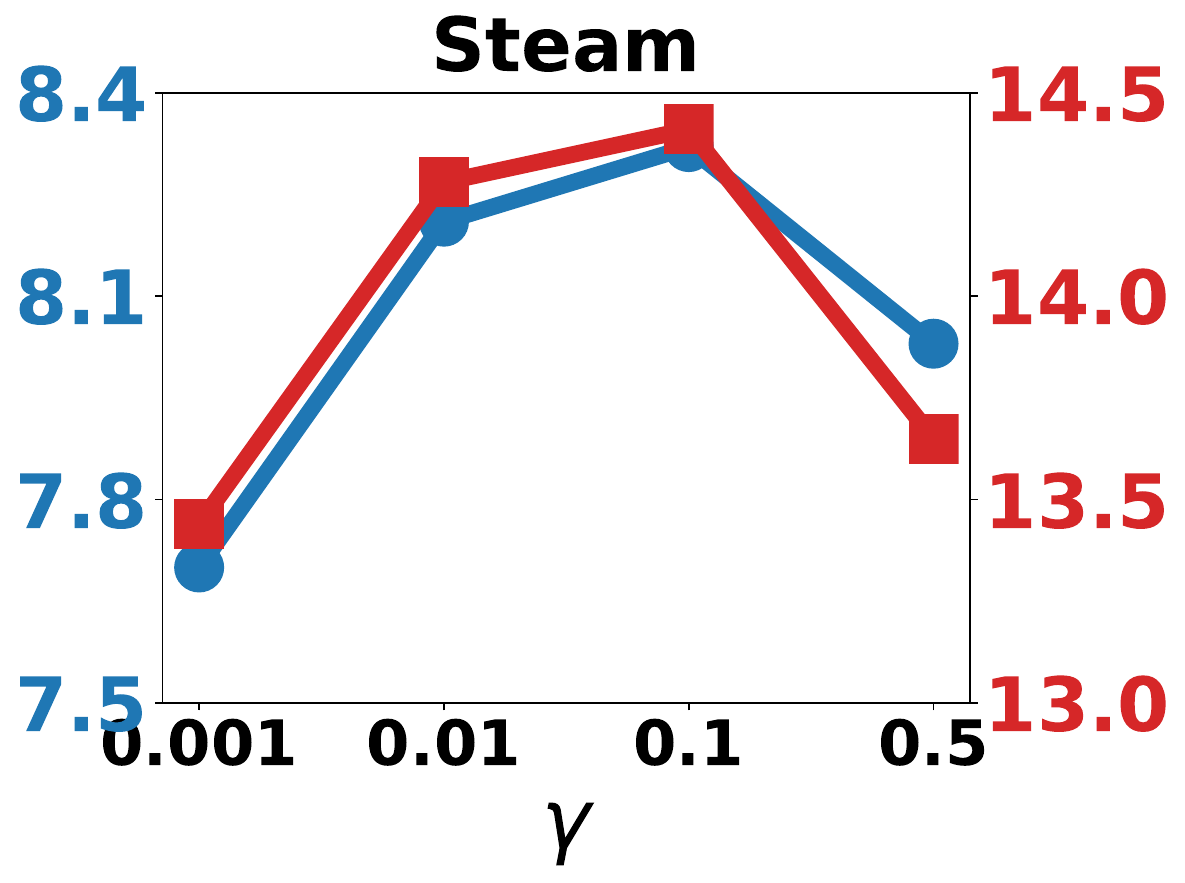}

    \includegraphics[width=0.3\linewidth]{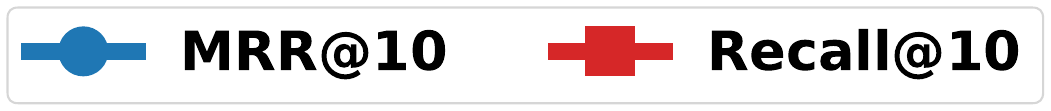}

    \caption{Hyper-parameter sensitivity analysis of $\gamma$ on three datasets}
    \label{fig:hyper}
\end{figure*}

\begin{figure*}[htbp]
    \centering
    \includegraphics[width=0.25\linewidth]{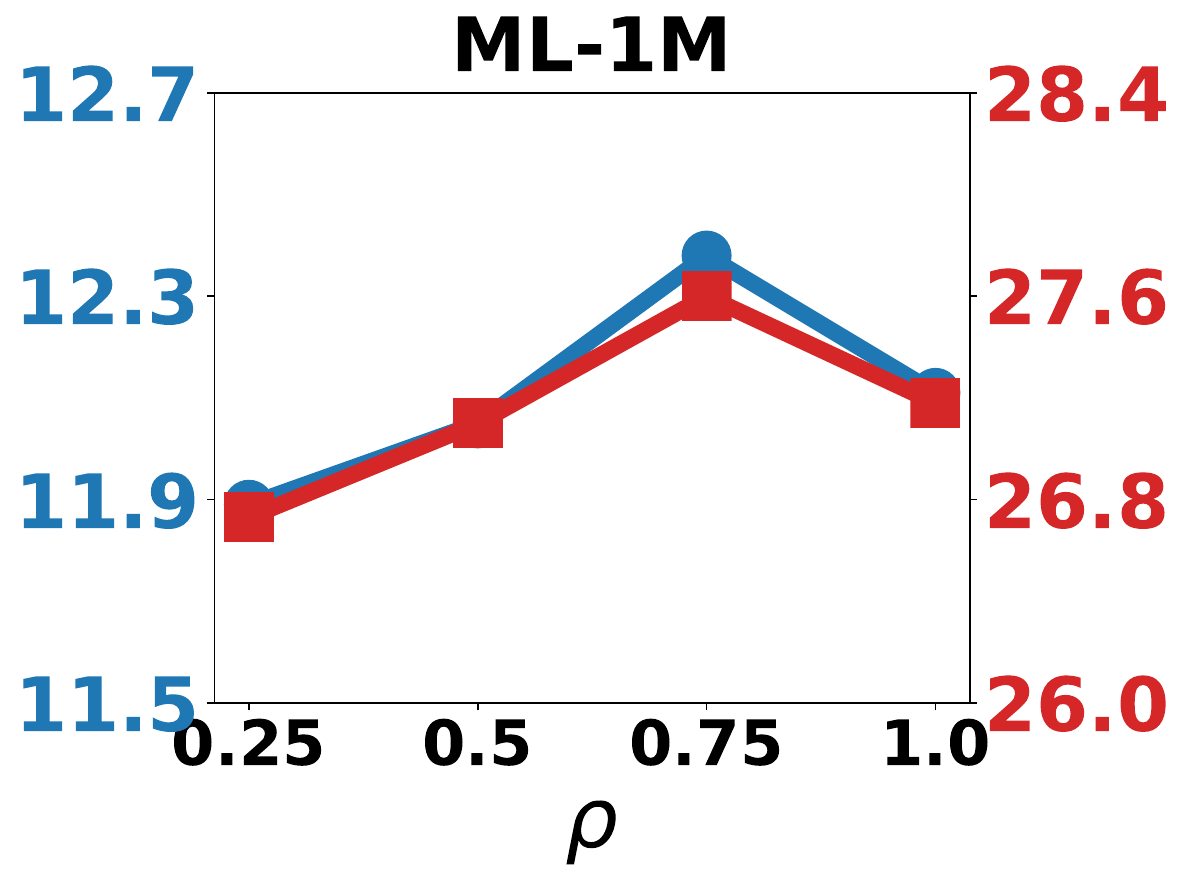}
    \includegraphics[width=0.25\linewidth]{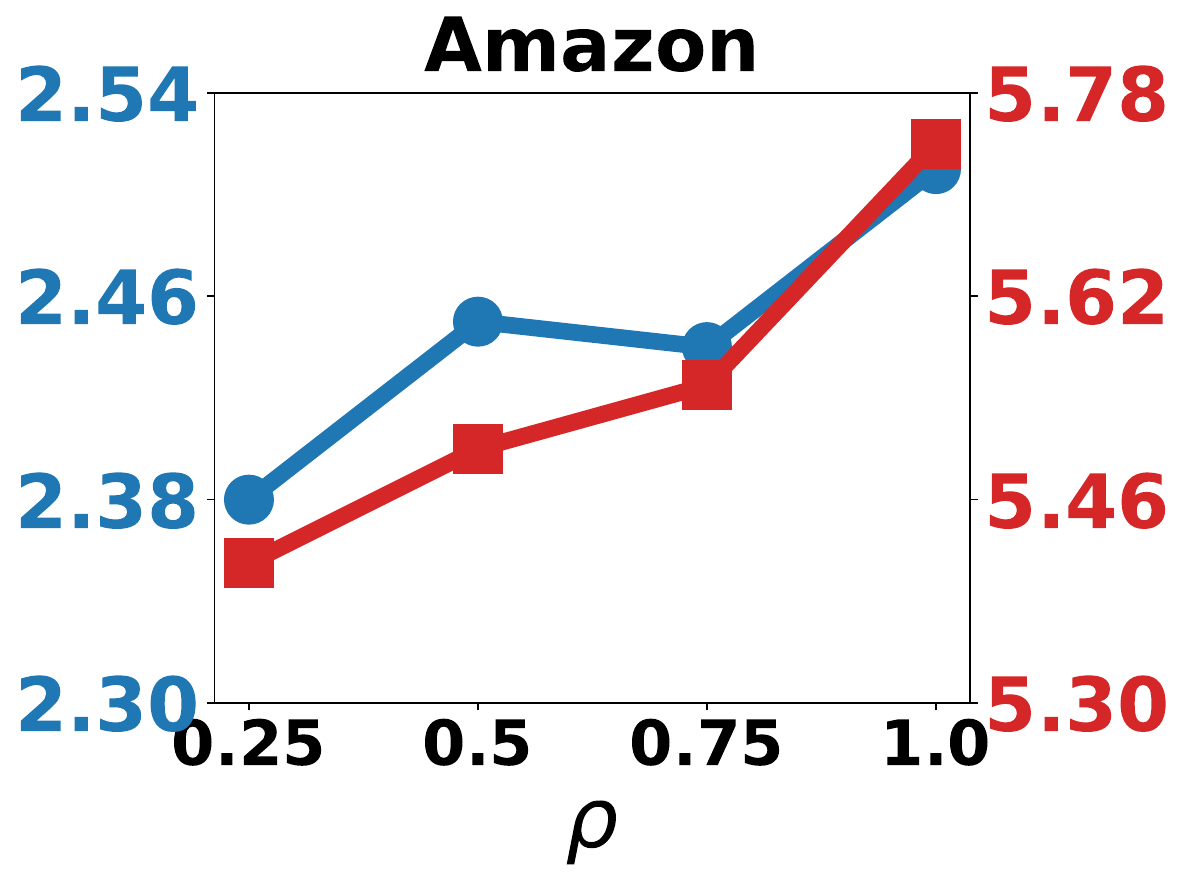}
    \includegraphics[width=0.25\linewidth]{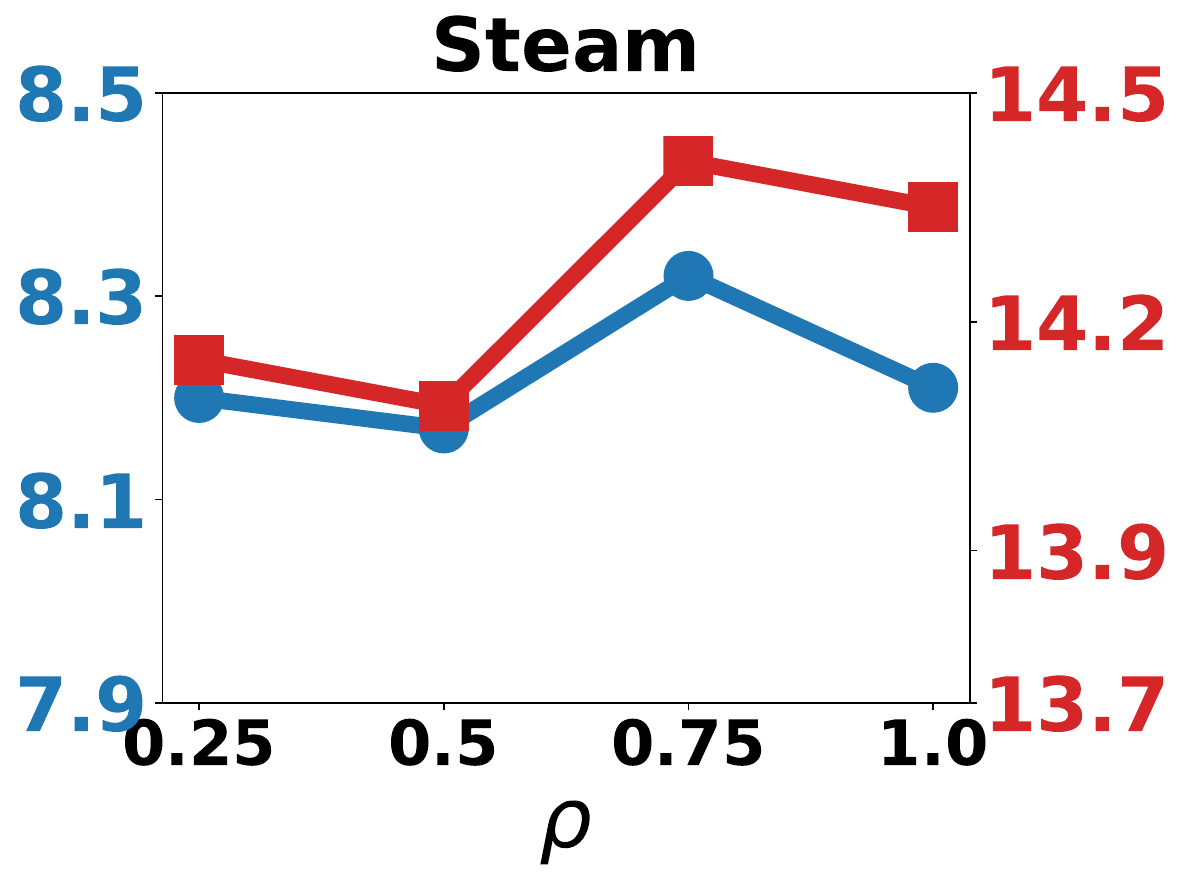}
    \includegraphics[width=0.3\linewidth]{figs/legend.pdf}
    \caption{Hyper-parameter sensitivity analysis of $\rho$ on three datasets}
    \label{fig:hypertho}
\end{figure*}

\begin{figure*}[htbp]
    \centering
    \includegraphics[width=0.25\linewidth]{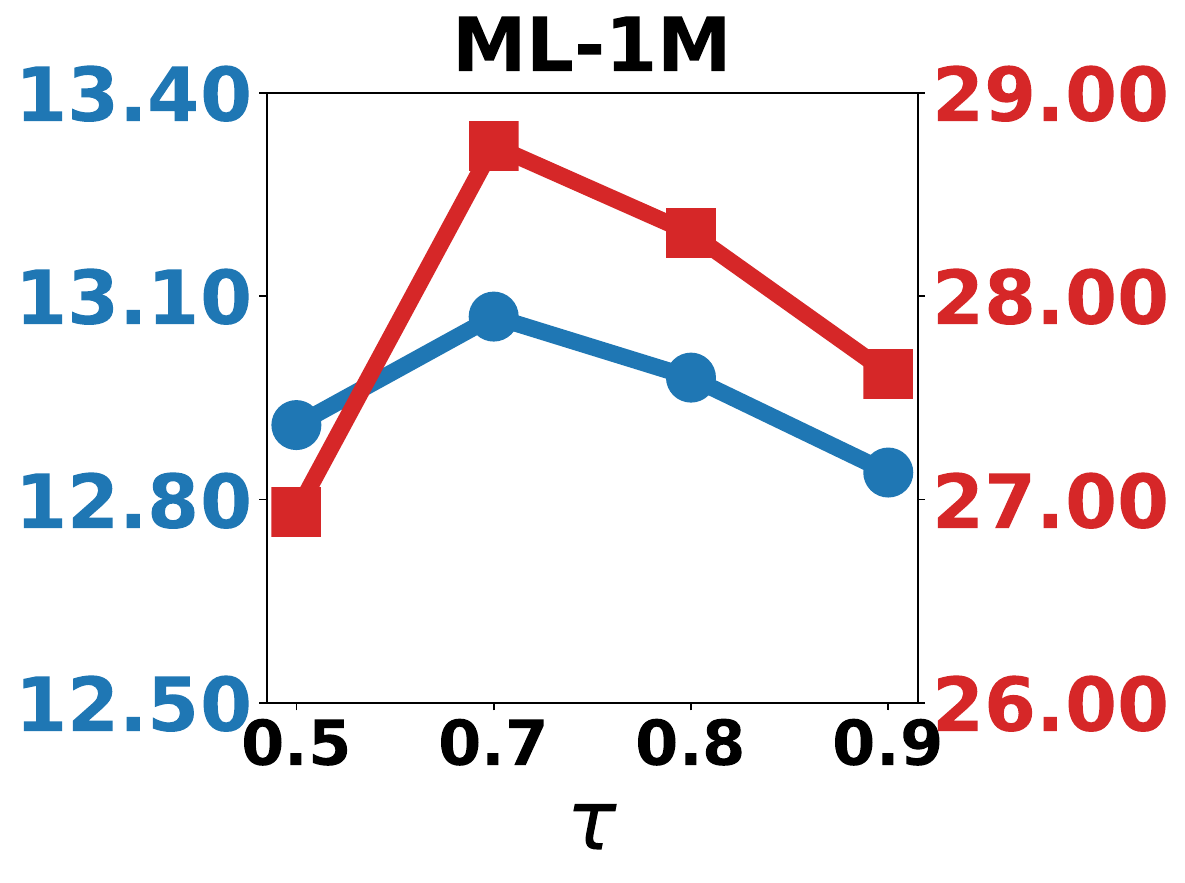}
    \includegraphics[width=0.25\linewidth]{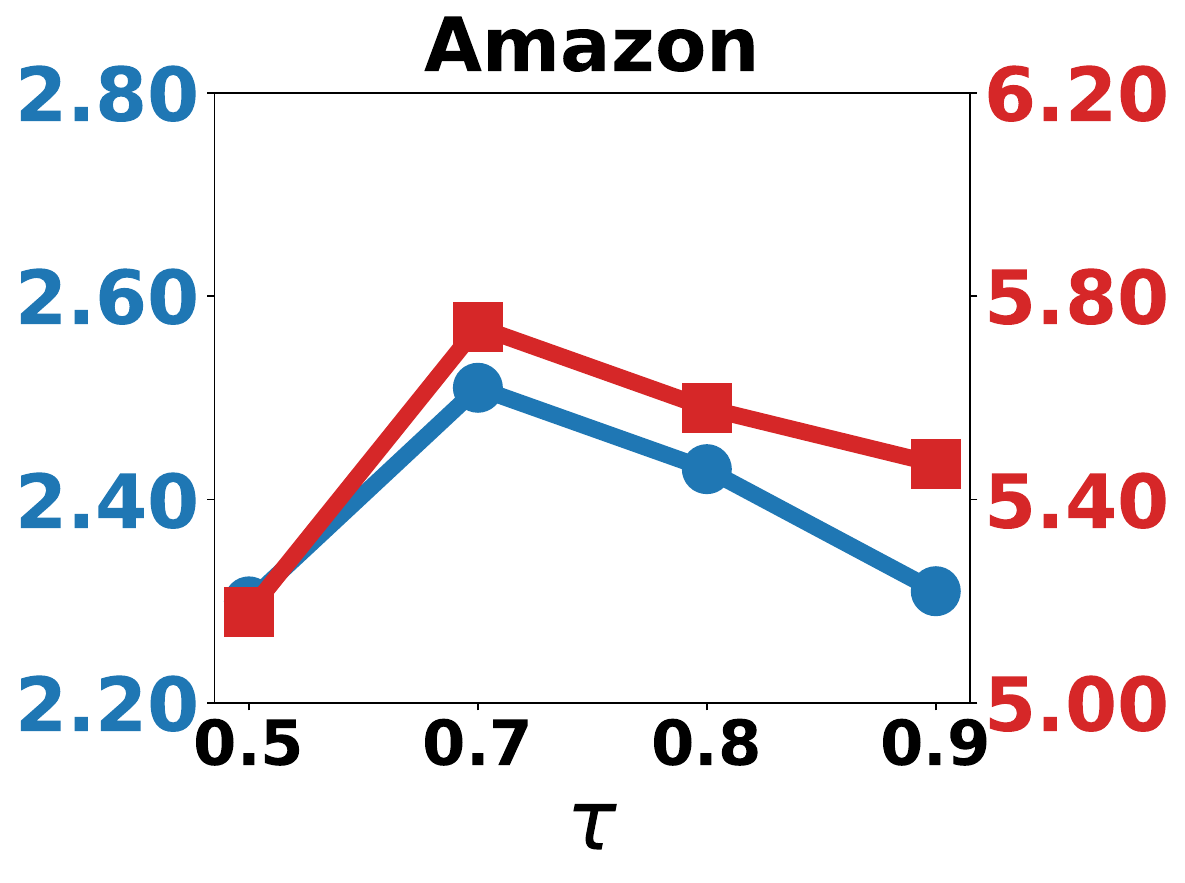}
    \includegraphics[width=0.25\linewidth]{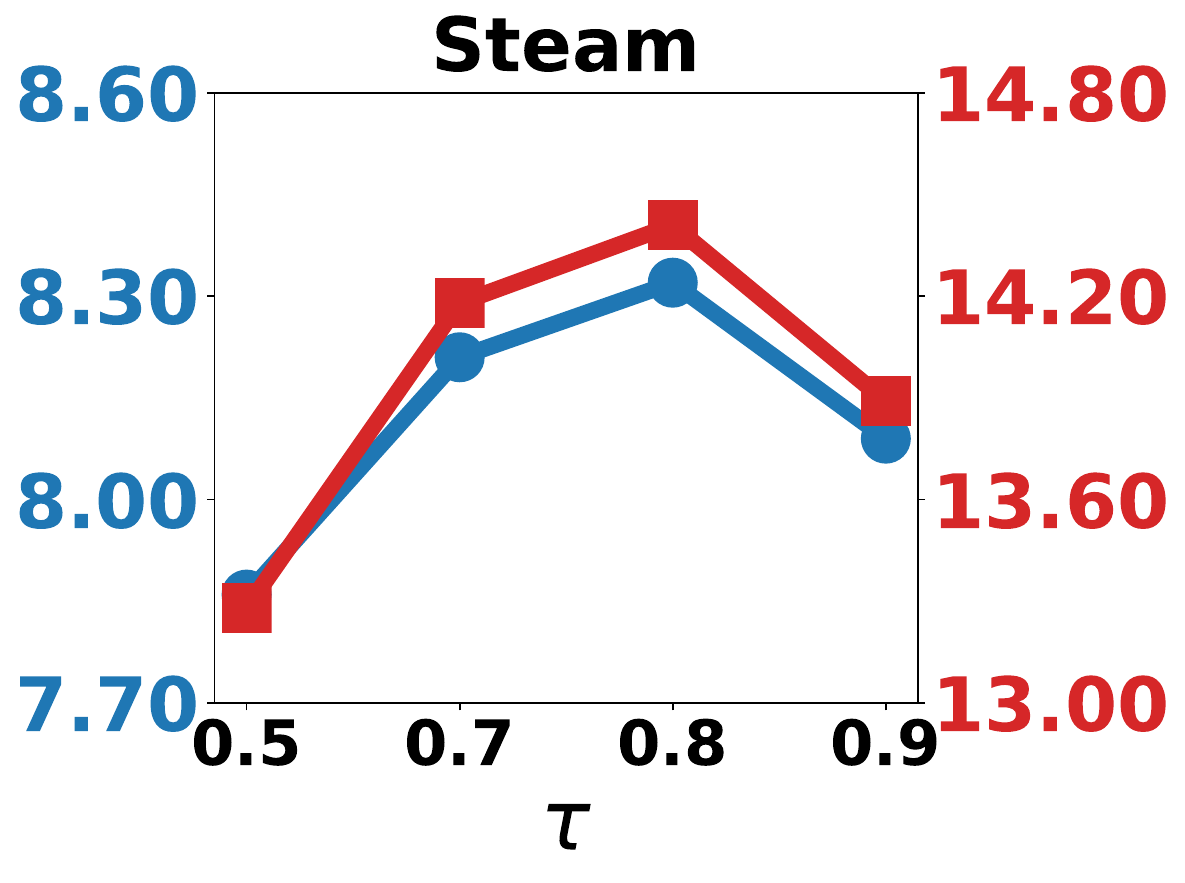}
    \includegraphics[width=0.3\linewidth]{figs/legend.pdf}
    \caption{Hyper-parameter sensitivity analysis of $\tau$ on three datasets}
    \label{fig:hypertau}
\end{figure*}

We examine the sensitivity of our model to three key hyperparameters:
the contrastive loss weight~$\gamma$, the compression ratio~$\rho$, and energy threshold $\tau$.
For each hyperparameter, we vary its value while holding all others
fixed at their optimal settings, and report Recall@10 and MRR@10
on ML-1M, Amazon, and Steam.
The results are presented in Figure~\ref{fig:hyper}, Figure~\ref{fig:hypertho}, and Figure~\ref{fig:hypertau}.

The parameter $\gamma$ governs the contribution of the contrastive
objective to the overall training loss.
As shown in the first row of Figure~\ref{fig:hyper}, performance
first improves and then declines as $\gamma$ increases from 0.001
to 0.5 across all three datasets, with the optimal value at
$\gamma = 0.01$ on ML-1M and $\gamma = 0.1$ on Amazon and Steam.
When $\gamma$ is too small, the cross-path alignment signal between
the Sequential Inference Encoder and the Hierarchical Interest Encoder
is insufficient to regularize the geometry of $\mathbf{u}$,
resulting in under-differentiated user representations.
Conversely, an excessively large $\gamma$ causes the contrastive term
to dominate the recommendation objective, impairing the sequential
modeling capacity of the SIE and degrading ranking quality.
These results confirm that $\mathcal{L}_{\text{CL}}$ functions as
an effective auxiliary regularizer rather than the primary training
signal, and that its weight requires moderate tuning to balance
cross-path alignment against next-item prediction fidelity.


The compression ratio $\rho$ governs the number of active interest
slots relative to the segment count~$K$ produced by the frozen
Soft-TPP module, via $U_{\text{eff}} = \min(\lceil \rho K \rceil,
U_{\max})$.
As shown in Figure~\ref{fig:hypertho}, the optimal $\rho$ varies across
datasets in a manner consistent with differences in sequence length
and the degree of interest recurrence within each domain.

On ML-1M, users accumulate substantially longer interaction histories (average length 165.50, compared to 23.61 on Amazon and 44.99 on Steam), which provides more positional opportunities for boundary events to accumulate, causing the Soft-TPP to produce a larger number of boundary-delimited segments $K$ per user. However, a large $K$ does not imply a proportionally larger number of distinct interests: given the narrow category vocabulary (18 categories), the probability of genuine cross-segment intent divergence is limited, meaning many segments are likely manifestations of the same underlying intent type rather than truly distinct behavioral phases.
While on the Amazon, sequences are considerably shorter, so the TPP tends to produce fewer segments per user. With a smaller baseline~$K$, interest recurrence is less prevalent,
and the segment-to-interest ratio is closer to one-to-one.
Compressing the slot budget in this regime ($\rho < 1.0$) would
prematurely merge segments that correspond to genuinely distinct
behavioral phases, discarding useful diversity in the multi-interest representation.
Retaining all slots ($\rho = 1.0$) therefore preserves the full
discriminative structure of~$\mathbf{z}$ without introducing the
redundancy that would arise from overly long histories.

Finally, the energy threshold $\tau$ controls the sensitivity of the Soft-TPP boundary detector: a smaller $\tau$ requires more severe energy dissipation before declaring a new segment, yielding coarser segmentation, while a larger $\tau$ triggers boundaries more readily, producing finer-grained segments. As shown in Figure~\ref{fig:hypertau}, the optimal $\tau$ varies across datasets in a manner consistent with their temporal characteristics. On ML-1M, where users frequently rate multiple movies within the same day, a large proportion of consecutive interactions carry near-zero time-gaps, causing the energy retention ratio $r_i$ to cluster near its upper bound. A higher threshold ($\tau = 0.8$) is therefore required to trigger meaningful boundaries at positions where genuine intent transitions occur, while smaller values fail to distinguish true phase changes from same-session continuations. On Amazon, the sparser and more category-diverse interaction logs produce larger inter-event time-gaps and more frequent categorical divergence, driving $r_i$ to lower values on average; the optimal threshold ($\tau = 0.7$) is thus reached earlier, and increasing $\tau$ beyond this point introduces spurious boundaries that fragment coherent browsing sessions. Steam exhibits a pattern similar to ML-1M, with the optimal threshold at $\tau = 0.8$, consistent with its domain-concentrated nature where users engage with games over extended periods within stable interest phases despite the long overall sequence duration. 

\subsection{ Computational Efficiency (RQ5)}\label{sec:exp6}

We compare the computational cost of S2-CAR against four representative 
baselines on the Amazon dataset, measuring per-epoch actual training time, peak 
GPU memory, and epochs to convergence. Results on the other two datasets follow the same relative ordering and are omitted for brevity.

\begin{table}[htbp]
\centering
\caption{Computational efficiency comparison on Amazon. Training time per epoch 
and peak GPU memory are measured on a single NVIDIA RTX~A5000 GPU. 
$^\dagger$ denotes the TPP training cost, 
which is a one-time overhead and excluded from the per-epoch timing.}
\label{tab:efficiency}
\resizebox{0.8\textwidth}{!}{%
\begin{tabular}{lccc}
\toprule[1pt]
\textbf{Method} & \textbf{Time/Epoch (s)} & \textbf{GPU Mem (MB)} & \textbf{Epochs to Converge} \\
\midrule
SASRec    & 58 & 6,663 & 13 \\
CL4SRec   & 231 & 13,955 & 10 \\
ICSRec    & 176 &  15,970 & 9 \\
BASRec    & 71 & 11,083 &  26 \\
\midrule
\textbf{S2-CAR}    & 144 & 11,375 & 7 \\
\textit{TPP pre-train}$^\dagger$ & 9 & 1,162 & 5 \\
\bottomrule[1pt]
\end{tabular}%
}
\end{table}

Table~\ref{tab:efficiency} reports the per-epoch training time, peak GPU memory,
and epochs to convergence for the proposed method and baselines on the Amazon dataset. S2-CAR's
per-epoch time (144~s) is higher than BASRec (71~s) and SASRec (58~s) due to
the additional hierarchical encoder, but it converges in only 7 epochs, bringing the total training time to 1,008 seconds, competitive with BASRec (71~s $\times$ 26 epochs = 1,846~s total) and ICSRec (176~s $\times$ 9 epochs = 1,584~s total).
In terms of memory, S2-CAR (11,375~MB) is comparable to BASRec (11,083~MB) and
considerably more efficient than ICSRec (15,970~MB).
The Soft-TPP pre-training adds only 45 seconds in total (9~s $\times$ 5 epochs),
which is negligible relative to the main-model training cost. The
complexity-adaptive slot gating ($U_{\text{eff}} = \min(\lceil\rho K\rceil, U_{\max})$)
selects a compact active subset from the $U_{\max}$ candidate interest vectors.
In the current implementation, candidate interest vectors are still computed
from $U_{\max}$ queries, so the main cross-attention cost remains
$\mathcal{O}(U_{\max} \cdot K \cdot d)$. The adaptive selection reduces the
subsequent pooling and diversity-regularization computation over active
interest vectors, and the entire HIE branch is discarded at serving time.

\subsection{Middle Information Utilization (RQ6)}\label{sec:exp4}
{\color{black}

We investigate whether S2-CAR improves the utilization of middle-sequence information by evaluating its reliance on different parts of the interaction history and its robustness to boundary-level perturbations. To this end, we conduct two complementary experiments: (1) an item deletion analysis that measures the contribution of different temporal segments, and (2) a boundary noise injection experiment that probes whether segmentation can isolate cross-segment interference.

Prior analysis (\Cref{fig:right}) reveals a clear asymmetry in SASRec: removing items from the \emph{middle} segment causes little or no performance degradation, and may even lead to slight improvements for longer sequences, whereas removing items from the \emph{last} segment consistently degrades performance. This indicates that existing models largely under-utilize middle-segment history and over-rely on recent interactions.

\Cref{fig:truncation_rq6} extends the same item deletion protocol to S2-CAR across all three benchmark datasets. In contrast to SASRec, removing items from the middle segment consistently leads to performance degradation across all length groups (orange bars, $\Delta\text{Recall@10} < 0$ throughout, where $\Delta\text{Recall@10}$ denotes the deleted-sequence score minus the original score). This result shows that S2-CAR actively relies on middle-sequence information rather than treating it as redundant. Meanwhile, removing items from the last segment (black bars) also degrades performance, confirming that recent interactions remain important. Together, these results indicate that S2-CAR achieves a more balanced utilization of both short-term and long-term signals.

\begin{figure}[htbp]
    \centering
    \begin{subfigure}[b]{0.32\textwidth}
        \centering
        \includegraphics[width=\textwidth]{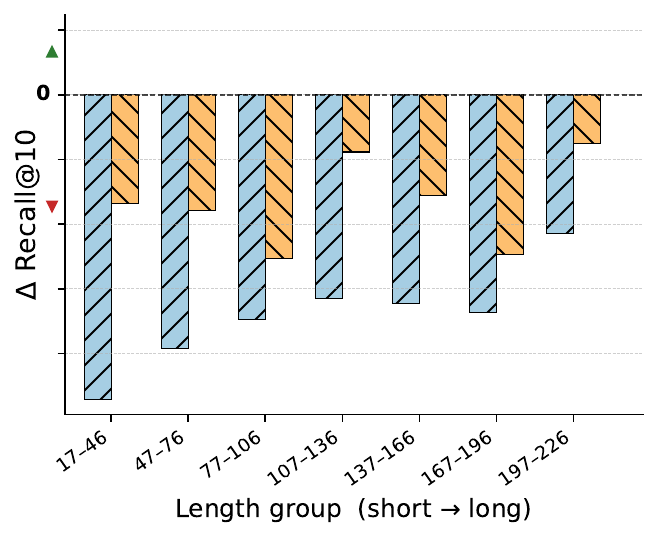}
        \caption{ML-1M}
        \label{fig:trunc_ml1m}
    \end{subfigure}
    \hfill
    \begin{subfigure}[b]{0.32\textwidth}
        \centering
        \includegraphics[width=\textwidth]{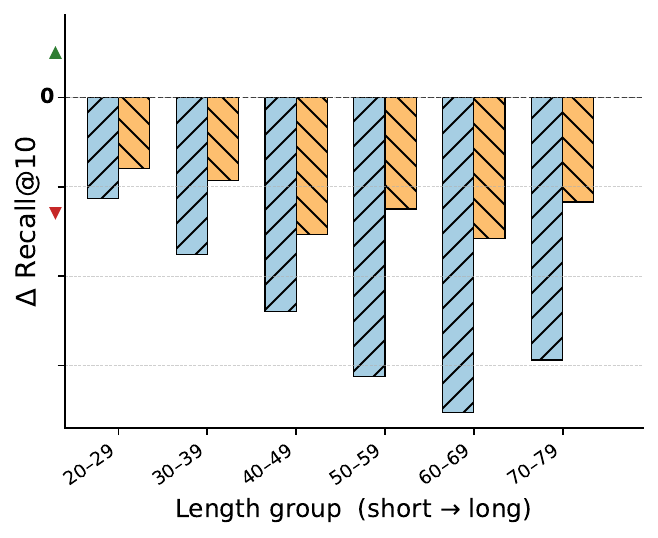}
        \caption{Amazon (length $\geq 20$)}
        \label{fig:trunc_amazon}
    \end{subfigure}
    \hfill
    \begin{subfigure}[b]{0.32\textwidth}
        \centering
        \includegraphics[width=\textwidth]{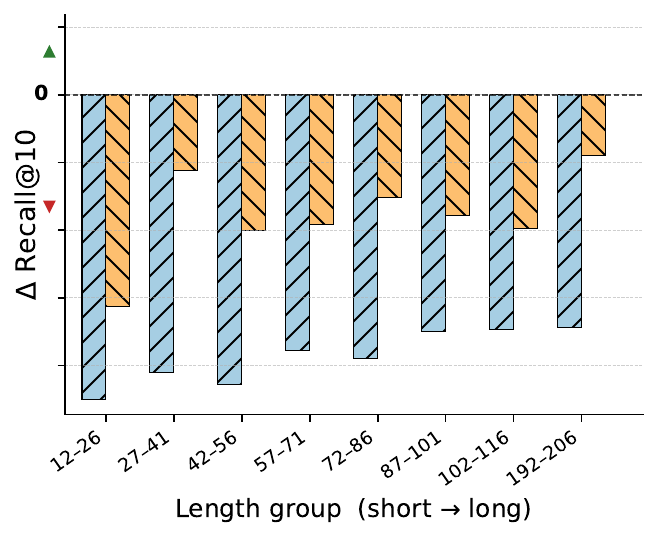}
        \caption{Steam}
        \label{fig:trunc_steam}
    \end{subfigure}
    \\[4pt]
    \includegraphics[width=0.55\textwidth]{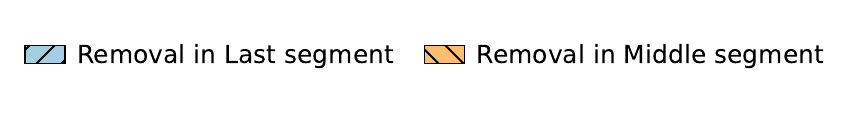}
    \caption{\color{black}Item deletion analysis for S2-CAR across all three datasets
    ($\Delta$\,Recall@10). Half the items in the \textit{last segment}
    (short-term proxy, black) or \textit{middle segment} (long-term
    proxy, orange) are randomly removed; bars below zero indicate
    performance degradation relative to the original sequence.}
    \label{fig:truncation_rq6}
\end{figure}

To further understand the underlying mechanism, we examine whether this improvement is driven by accurate intent segmentation. If segmentation correctly identifies intent boundaries, noise introduced at segment boundaries should be contained locally rather than affecting representations that capture longer-range dependencies. To test this, for each user sequence we insert a randomly sampled out-of-category item at a random position within either the \textit{last} temporal segment (\textbf{Tail Injection}) or the \textit{first} temporal segment (\textbf{Head Injection}), where segments are defined by the standard 48-hour inactivity threshold~\citep{chen2022time}. We deliberately adopt this fixed heuristic rather than TPP-derived boundaries to ensure that the injected noise is independent of S2-CAR's own segmentation, avoiding confounding effects in cross-model comparison. All models are retrained from scratch on the perturbed sequences under identical settings, and the performance change is reported as $(\text{perturbed} - \text{original}) / \text{original} \times 100\%$; because the reported values are negative under performance degradation, values closer to zero indicate greater robustness.

As shown in~\Cref{tab:rq4}, S2-CAR reports values closest to zero in most settings and remains competitive in the few cases where TLSRec is slightly closer to zero, such as Amazon R@5. Overall, the results suggest that boundary-level noise is less likely to dominate S2-CAR's sequence representation. In contrast, SASRec suffers the largest degradation, reflecting its strong reliance on positional bias. CL4SRec shows moderate robustness due to stochastic augmentations that reduce dependence on specific positions, while ICSRec benefits from intent-level clustering but still allows noise to affect the target intent representation in the absence of explicit boundary detection. TLSRec exhibits moderate robustness, consistently outperforming SASRec and CL4SRec, as its session-level modeling partially captures temporal structure. However, without explicit boundary detection, it remains susceptible to cross-segment noise in most metrics, leading to greater degradation than S2-CAR overall.

\begin{table}[htbp]
\centering
\caption{Relative performance degradation (\%) under boundary noise injection.}
\label{tab:rq4}
\resizebox{0.9\textwidth}{!}{%
\begin{tabular}{ll|cccc|c}
\toprule
\textbf{Dataset} & \textbf{Metric}
    & SASRec (\%) $\downarrow$ & CL4SRec (\%) $\downarrow$ & ICSRec (\%) $\downarrow$ & TLSRec (\%) $\downarrow$ & \textbf{S2-CAR (\%) $\downarrow$} \\
\midrule
\multirow{4}{*}{\textbf{ML-1M}}
 & R@5   & $-$9.83 & $-$6.41 & $-$4.72 & $-$3.89 & \textbf{$-$1.40} \\
 & R@10  & $-$9.12 & $-$5.87 & $-$4.31 & $-$4.17 & \textbf{$-$0.19} \\
 & MRR@5  & $-$10.24 & $-$6.93 & $-$5.01 & $-$4.52 & \textbf{$-$1.06} \\
 & MRR@10 & $-$10.57 & $-$7.12 & $-$4.88 & $-$4.81 & \textbf{$-$1.21} \\
\midrule
\multirow{4}{*}{\textbf{Amazon}}
 & R@5   & $-$7.41 & $-$5.23 & $-$4.18 & \textbf{$-$3.14} & $-$3.37 \\
 & R@10  & $-$6.63 & $-$4.97 & $-$3.85 & $-$3.95 & \textbf{$-$3.42} \\
 & MRR@5  & $-$8.62 & $-$5.64 & $-$4.53 & $-$4.17 & \textbf{$-$3.23} \\
 & MRR@10 & $-$8.21 & $-$5.38 & $-$4.29 & $-$4.75 & \textbf{$-$3.22} \\
\midrule
\multirow{4}{*}{\textbf{Steam}}
 & R@5   & $-$8.74 & $-$5.91 & $-$4.43 & $-$4.62 & \textbf{$-$1.70} \\
 & R@10  & $-$7.93 & $-$5.42 & $-$3.97 & $-$4.14 & \textbf{$-$0.37} \\
 & MRR@5  & $-$9.31 & $-$6.24 & $-$4.61 & $-$3.73 & \textbf{$-$1.62} \\
 & MRR@10 & $-$8.87 & $-$5.83 & $-$4.37 & $-$4.07 & \textbf{$-$1.29} \\
\bottomrule
\end{tabular}%
}
\end{table}
}

\section{Discussion}
{\color{black}
\subsection{Theoretical Implications}
S2-CAR advances the theoretical understanding of user intent modeling in sequential recommendation along three dimensions. First, by grounding intent segmentation in a Continuous-Time Temporal Point Process (TPP) framework, we provide a principled probabilistic alternative to heuristic time-gap or fixed-window partitioning strategies. The energy retention ratio $r_i$ derived from the Context-Aware Soft-TPP offers a theoretically motivated boundary criterion that adapts to the local categorical dynamics of each user, rather than imposing a globally uniform criterion. Second, the Segment-Count-Adaptive slot gating mechanism (Equation~\ref{eq:ueff}) formalizes the intuition that interest capacity should scale with behavioral complexity, addressing the uniform-slot limitation identified in prior multi-interest work~\citep{li2019multi,cen2020controllable}. Third, the decoupled training design, where the HIE provides contrastive supervision to the SIE without participating in inference, establishes a general principle for injecting structural regularization into autoregressive encoders without compromising their sequential fidelity.

\subsection{Practical Implications}
From a practical standpoint, S2-CAR offers several deployment-relevant advantages. The Soft-TPP segmentation module is pretrained once and frozen, adding only 45 seconds of one-time overhead (Table~\ref{tab:efficiency}), and functions as a plug-and-play component that consistently improves existing backbones, including augmentation-based (CoSeRec), graph-based (SURGE), and diffusion-based (DiffuRec) architectures, without modifying their internal design (Table~\ref{tab:backbone}). This modularity makes the proposed segmentation broadly applicable beyond the S2-CAR framework. Furthermore, S2-CAR converges in only 7 epochs despite its two-encoder architecture, resulting in a total training cost competitive with simpler baselines such as BASRec, while achieving substantially higher recommendation quality. These properties suggest that S2-CAR is suitable for real-world deployment scenarios where both accuracy and computational efficiency are critical.

\subsection{Differences from Existing Work}
Table~\ref{tab:feature_compare} summarizes how S2-CAR differs from existing methods along five modeling dimensions. Unlike prior hierarchical methods that rely on fixed temporal windows~\citep{chen2022time} or raw time-gap thresholds~\citep{ludewig2018evaluation}, S2-CAR infers segment boundaries from the continuous-time decay of latent intent energy, enabling task-specific segmentation without manually annotated boundary labels. Unlike multi-interest models that allocate a uniform number of interest slots across all users~\citep{li2019multi,choi2024multi}, S2-CAR dynamically adjusts slot capacity in proportion to each user's segment count via the compression ratio $\rho$, preventing both capacity waste and under-representation of diverse interests.

\subsection{Limitations and Future Work}
While S2-CAR achieves consistent improvements across three diverse benchmarks, several limitations merit discussion. First, the two-stage training requires that category information is available at both pretraining and inference time; datasets without category metadata would require an alternative decay conditioning mechanism, such as item-level semantic similarity or co-occurrence statistics. Second, the threshold hyperparameter $\tau$ requires dataset-specific tuning, as optimal values vary with the temporal density and categorical diversity of each domain (Section~\ref{sec:exp5}); the hyperparameter sensitivity analysis in Section~\ref{sec:exp5} shows that performance is relatively stable within a moderate range around the optimal value, but automating this selection remains an open direction. Third, while the Soft-TPP module is computationally lightweight ($\mathcal{O}(Nd)$ per sequence, pretrained once in 45 seconds), the auxiliary HIE branch increases the parameter count by approximately $1.7\times$ the parameter count of a single Transformer, increasing GPU memory requirements during training to around 11~GB (comparable to BASRec at 11~GB, but higher than SASRec at 6.6~GB, as reported in Table~\ref{tab:efficiency}); future work could explore more memory-efficient contrastive designs such as momentum encoders or asynchronous teacher updates. Finally, the current evaluation is limited to three benchmarks spanning movie, e-commerce, and gaming domains. Extending the evaluation to cross-domain, social media, or news recommendation settings, where intent dynamics may be triggered by different contextual signals, would further validate the generality of the proposed energy-based segmentation framework.
}
\section{Conclusion}

In this paper, we proposed S2-CAR, a Segmentation-Supervised Hierarchical Multi-Intent Network with Complexity-Adaptive Regularization for sequential recommendation. Our work is motivated by two complementary failure modes of existing approaches: the reliance on fixed temporal boundaries that misalign with true intent transitions, and the uniform allocation of interest capacity that fails to reflect the varying complexity of user behavior across sessions. To address these, S2-CAR introduces a Context-Aware Soft-TPP that models user intent as a continuously decaying latent energy state, triggering segment boundaries only when energy retention falls below a dataset-specific threshold, without manually annotated boundary labels. Our plug-in experiments show consistent improvements on representative sequential recommenders, suggesting that behaviorally adaptive boundary construction can serve as a useful structural prior for a broader class of sequence modeling frameworks.

\clearpage
\bibliographystyle{elsarticle-harv}
\bibliography{cas-refs}

\end{document}